\documentclass[%
 reprint,
superscriptaddress,
showpacs,
nofootinbib, nobibnotes, 
 amsmath,
 amssymb,
 aps,
 pre
]{revtex4-1}

\usepackage{blindtext}
\usepackage{dcolumn}
\usepackage{bm}
\usepackage{color, soul, colortbl} 
\usepackage[colorlinks=true,
						linkcolor=blue,
						urlcolor=blue,
						citecolor=blue,
						bookmarks=true,
						pdfborder={0 0 0}]{hyperref}
       
\usepackage[rgb]{xcolor}  

\usepackage{array,multirow}
\newcolumntype{+}{!{\vrule width 2pt}}

\usepackage{graphicx,epsfig}
\usepackage{subfigure}
\usepackage{amsmath}
\usepackage{xfrac}

\begin{document}

\title{Evaluating the impact of PrEP on HIV and gonorrhea\\on a networked population of female sex workers}

\author{Alba Bernini}
	\affiliation{Dipartimento di Elettronica, Informazione e Bioingegneria, Politecnico di Milano, Milan, Italy}
\author{Elodie Blouzard}
	\affiliation{Faculty of veterinary science, Utrecht university, The Netherlands}
\author{Alberto Bracci}
	\affiliation{Department of Applied Science and Technology (DISAT), Politecnico di Torino, Corso Duca degli Abruzzi 24, Torino, Italy}
\author{Pau Casanova}
	\affiliation{Grupo Interdisciplinar de Sistemas Complejos (GISC), Department of Mathematics, Carlos III University of Madrid, Legan\'es, Spain}
	\affiliation{Centro Nacional de Biotecnolog\'ia (CSIC), Darwin 3, 28049 Madrid, Spain.}
\author{Iacopo Iacopini}
    \affiliation{School of Mathematical Sciences, Queen Mary University of London, London E1 4NS, United Kingdom}
	\affiliation{Centre for Advanced Spatial Analysis, University College London, London, W1T 4TJ, United Kingdom}
\author{Benjamin Steinegger}
	\affiliation{Departament d'Enginyeria Inform\`atica i Matem\`atiques, Universitat Rovira i Virgili, E-43007 Tarragona, Spain}
\author{Andreia Sofia Teixeira}
	\affiliation{INESC-ID and Instituto Superior T\'ecnico, Universidade de Lisboa, Lisbon, Portugal}
	\affiliation{ATP-Group, IST-Taguspark, Porto Salvo, Portugal}
	\affiliation{LASIGE and Faculdade de Ci\^encias, Universidade de Lisboa, Lisbon, Portugal}
\author{Alberto Antonioni}
	\affiliation{Grupo Interdisciplinar de Sistemas Complejos (GISC), Department of Mathematics, Carlos III University of Madrid, Legan\'es, Spain}
\author{Eugenio Valdano}
	\email[E-mail: ]{eugenio.valdano@gmail.com}
	\affiliation{Center for Biomedical Modeling, The Semel Institute for Neuroscience and Human Behavior, David Geffen School of Medicine, 760 Westwood Plaza, University of California Los Angeles, Los Angeles, CA 90024, USA}

\date{\today}

\begin{abstract}
Sexual contacts are the main spreading route of HIV. This puts sex workers at higher risk of infection even in populations where HIV prevalence is moderate or low. Alongside condom use, Pre-Exposure Prophylaxis (PrEP) is an effective tool for sex workers to reduce their risk of HIV acquisition. However, PrEP provides no direct protection against sexually transmitted infections (STIs) other than HIV, unlike condoms. 
We use an empirical network of sexual contacts among female sex workers (FSWs) and clients to simulate the spread of HIV and gonorrhea.
We then investigate the effect of PrEP adoption and adherence, on both HIV and gonorrhea prevalence.
We also study the effect of a potential increase in condomless acts due to lowered risk perception with respect of the no-PrEP scenario (risk compensation).
We find that when HIV is the only disease circulating, PrEP is effective in reducing HIV prevalence, even with high risk compensation. Instead, the complex interplay between the two diseases shows that different levels of risk compensation require different intervention strategies.
Finally, we find that providing PrEP only to the most active FSWs is less effective than uniform PrEP adoption. Our work shows that the effects emerging from the complex interactions between these diseases and the available prophylactic measures need to be accounted for, to devise effective intervention strategies.
\end{abstract}

\maketitle

\section*{Introduction} 
Sex workers are disproportionately vulnerable to HIV, especially in low and middle-income countries~\cite{world2012prevention, baral2012burden}.
Violence, stigma, and punitive laws hinder their access to healthcare services~\cite{shannon_global_2015}.
In addition, scarcely available condoms, and low bargaining power with clients who refuse to use them, dramatically increase sex workers' risk of HIV acquisition~\cite{ghimire_reasons_2011, shannon2009structural, urada2012condom}. 


Pre-Exposure Prophylaxis (PrEP) represents an additional prevention tool, and consists in taking regularly a specific antiretroviral medication. It has the potential of protecting HIV-uninfected sex workers from acquiring HIV, especially if they cannot use condoms regularly.
Furthermore, access to PrEP helps sex workers have personal control over their health, with positive implications beyond HIV prophylaxis~\cite{eakle_i_2019}.
High efficacy in reducing the risk of HIV acquisition in highly adherent individuals has been extensively investigated and reported~\cite{flash_pre-exposure_2017, fonner2016effectiveness}. Notwithstanding, reaching the required adherence among sex workers might prove difficult, if the roll-out of PrEP is not supported by formative activities~\cite{galea2011acceptability, reza2016prioritizing, eakle2018designing, ortblad2018acceptability, mboup2018early, eakle_i_2019, bazzi2019pr}. 
In addition, PrEP might influence the frequency at which sex workers use condoms, although no definite evidence in this sense exists.
The potential cause of this is risk compensation, i.e. an increase in risky behavior sparked by a decrease in perceived risk~\cite{grant_when_2017, cassell2006risk, blumenthal2014risk}.
If PrEP was to lead sex workers towards riskier sexual behavior, it would increase their probability of acquiring other Sexually Transmitted Infections (STIs), such as gonorrhea, chlamydia, syphilis~\cite{nguyen2018incidenceSTIbeforeafterprep}. This in turn can increase the risk of acquiring HIV~\cite{kalichman2011prevalenceSTIcoinfinHIVpos}, especially if the adherence to PrEP is not optimal. Consequently, concerns about risk compensation remain a consistent barrier to PrEP provision among health care providers~\cite{mugwanya2013sexual,pilgrim2018provider}.

One of the most concerning HIV co-infections is gonorrhea (NG), caused by the bacteria \textit{Neisseria gonorrhoeae}. It occurs in approximately 9.5\% of people living with HIV (PLHIV)~\cite{kalichman2011prevalenceSTIcoinfinHIVpos}, and it has been showing increasing antimicrobial resistance~\cite{unemo2014AMRgono}. 
The impact of PrEP on gonorrhea is difficult to assess. On the one hand, lower condom use could enhance the probability of contracting gonorrhea. On the other hand, the periodical screenings that are associated with regular PrEP use may entail a better detection of gonorrhea (especially in its asymptomatic form), and lead to a decrease in its prevalence~\cite{jenness2017incidencegonochlamyinHIVpopMSM}.
Summing up, this is a complex and still open issue.

Mathematical models can help shed light on the complex interaction among HIV, gonorrhea, and prophylaxis, and inform optimal strategies for PrEP roll-out. Previous modeling works have focused on predicting the reduction in new HIV acquisitions, on the cost-effectiveness of PrEP in different countries, focusing on different key populations \cite{zhong_modeling_2018,pretorius2010evaluating}. However, little attention has been payed to the interaction between HIV and other STIs \cite{mushayabasa2011modeling}. In addition, they have assumed that sex workers and clients mix homogeneously, at odds with known results highlighting the role of complex contact patterns in shaping epidemic processes~\cite{moreno2002epidemic,salathe2010high,Barrat2008b,gauvin2013activity,barrat2008dynamical}. Specifically, heterogeneities in the number of contacts, and in their occurrence in time, are known to impact the performance of immunization strategies~\cite{holme2017cost} and of interaction between different \cite{Schiffman2003} pathogens~\cite{Schiffman2003,Cohen16302,poletto2015characterising,chen2017fundamental,Pinotti2018}. 

\medskip

In this study, we simulate the concurrent spread of HIV and gonorrhea over a network of real sexual contacts among female sex workers (FSWs) and their male clients, collected in Brazil over the course of six years~\cite{rocha2010information,rocha2011simulated}.
We investigate the impact of PrEP as a prophylactic measure additional to condoms, and the role of risk compensation. We do that both by assuming the spread of HIV only, and both HIV and gonorrhea.
Finally, we study the impact of nonuniform PrEP adoption on both HIV and gonorrhea prevalence.



\section*{Materials and methods}

\subsection*{Dataset}\label{data}
\noindent The anonymized dataset is obtained from a Brazilian web community of posts about self-reported sexual encounters between female sex workers and male clients. Each encounter is represented as a link of a bipartite network composed of $N^{[F]}=5 965$ female and $N^{[M]}=8 818$ male nodes. We should notice that $33 875$ out of the $44 088$ reported contacts involve a unique pair of nodes. By taking advantage of this longitudinal nature of the dataset, we consider daily snapshots of the interactions, resulting in a temporal network~\cite{holme2013temporal} that spans over $1 233$ days (we remove the first $1 000$ days of the original dataset). Repeated interactions among the same pair of nodes within the same day are not considered.

\begin{figure}[t]
	\centering
	\includegraphics[width=0.48\textwidth]{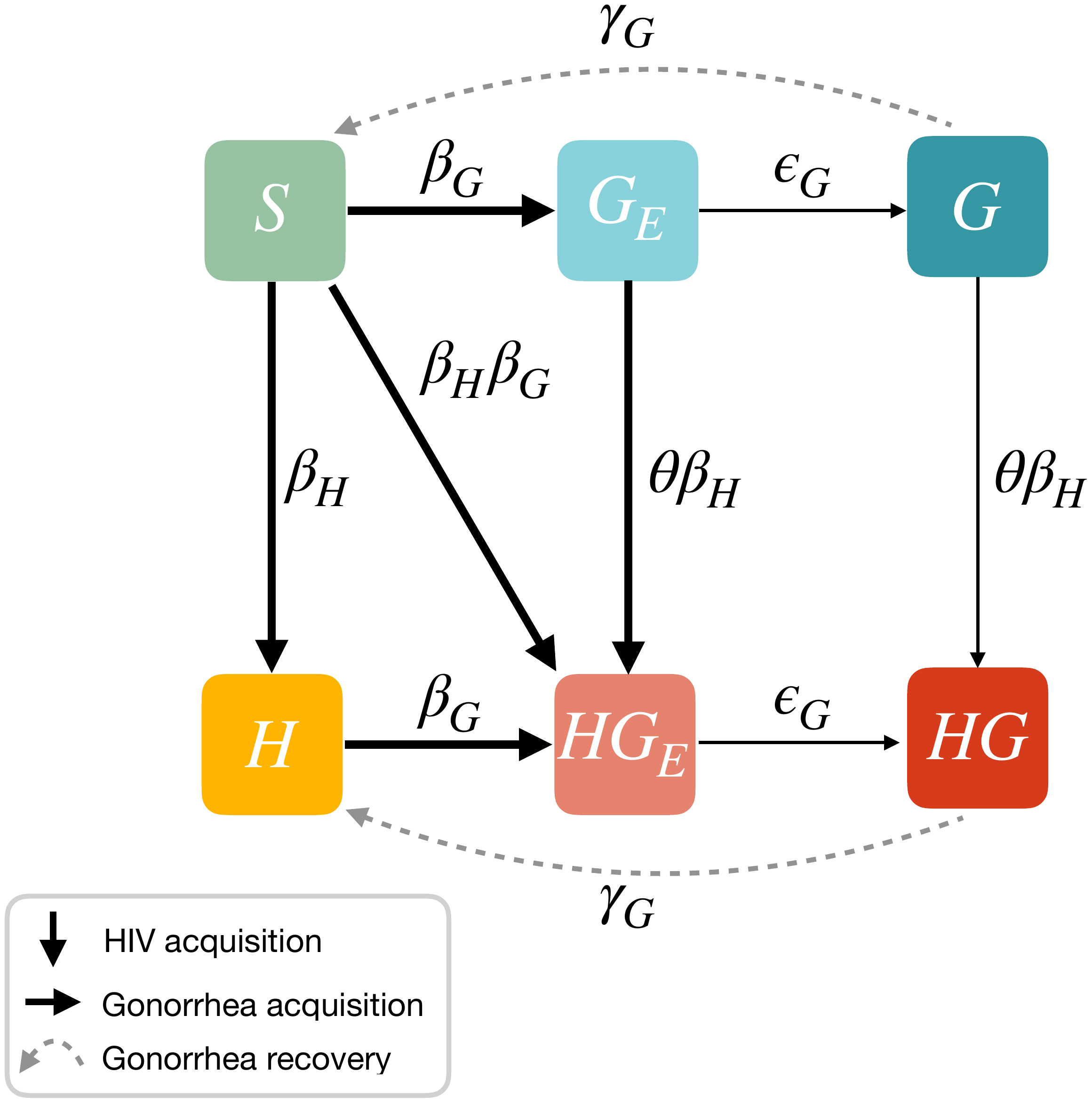}
	\caption{\textbf{The model.} 
		Individuals can belong to one of the six compartments. Susceptible individuals ($S$) can acquire either HIV (H, $\downarrow$) or gonorrhea (G, $\rightarrow$) after a heterosexual contact with an infected person. Once infected by gonorrhea, individuals become gonorrhea-exposed ($G_E$), i.e. infected by gonorrhea but not able to infect others yet. After a certain latency period, individuals can become gonorrhea-infectious ($G$) and thus able to transmit the disease. Once infected by HIV, individuals are immediately HIV-infectious ($H$), i.e. able to infect other susceptible persons. At the same time, HIV-infectious can also become gonorrhea-exposed ($HG_E$). Again, after a certain latency period individuals HIV-infectious can turn gonorrhea-infectious ($HG$). While recovery from gonorrhea is possible, HIV-infected individuals cannot become susceptible.}
	\label{fig:model_diagram}
\end{figure}

\subsection*{Epidemic model}\label{model}
\noindent We define a stochastic epidemic model to simulate the spread of HIV and gonorrhea over the networked population. In particular, we use a compartmental model (see Fig.~\ref{fig:model_diagram}) in which individuals are divided into six different compartments, according to their status with respect to the considered diseases: 
\begin{itemize}
	\item $S$: susceptible individuals who can acquire both HIV and gonorrhea by means of a sexual contact; 
	\item $G_E$: exposed individuals who have acquired gonorrhea but are not yet able to transmit it; 
	\item $G$: gonorrhea-infectious individuals who have gonorrhea and can transmit it; 
	\item $H$: HIV-infectious individuals who have HIV and can transmit it;
	\item $HG_E$: HIV-infectious individuals who have also been exposed to gonorrhea, but can only transmit HIV;
	\item $HG$: infectious individuals who can transmit both HIV and gonorrhea.
\end{itemize}

\begin{table}
	\setlength\extrarowheight{3pt}
	\centering
	\begin{tabular}{|c|r|c|c|c|}
		\cline{3-5}
		\multicolumn{1}{c}{}&	&{\bf Parameter} & {\bf Value} & {\bf Source}\\
		\hline
		\parbox[t]{2mm}{\multirow{5}{*}{\rotatebox[origin=c]{90}{{\bf gonorrhea}}}}
		&Infection probability                       & $\beta^{M\rightarrow F}_G$    & 0.7       &  \cite{kretzschmar_modeling_1996}\\
		\cline{2-5}
		&Infection probability    & $\beta^{F\rightarrow M}_G$    & 0.3      &\cite{kretzschmar_modeling_1996} \\
		\cline{2-5}
		&Latency period                                     & $1/\epsilon^M_G$              & 3.5 days  &\cite{kretzschmar_modeling_1996}  \\
		\cline{2-5}
		&Latency period                  & $1/\epsilon^F_G$              & 10 days   &\cite{kretzschmar_modeling_1996}  \\
		\cline{2-5}
		&Infectious period                          & $1/\gamma_{G}$            & 16 months  &\cite{kretzschmar_modeling_1996} \\
		\hline

		\parbox[t]{2mm}{\multirow{3}{*}{\rotatebox[origin=c]{90}{{\bf HIV}}}}
		&Infection probability                       & $\beta^{F\rightarrow M}_H$    & 0.0057    &\cite{bekker_combination_2015} \\
		\cline{2-5}
		& Infection probability    & $\beta^{M\rightarrow F}_H$    & 0.029     &\cite{bekker_combination_2015}  \\
		\cline{2-5}
		&Increase in HIV transm.                & $\theta$                      & 10        &\cite{chesson_sexually_2000}  \\
		\cline{1-5}
		\cline{2-5}
		\multicolumn{1}{c|}{}
		&Efficacy of condoms      & $\eta_{cond}$                 & 0.8       &\cite{mukandavire_comparing_2016}  \\
		\cline{2-5}
		\multicolumn{1}{c|}{}
		&Efficacy of PrEP                            & $\eta_{PrEP}$                 & 0.7       &\cite{mukandavire_comparing_2016} \\
		\cline{2-5}
		\cline{2-5}
		\multicolumn{1}{c|}{}
		&Prob. condom adoption    & $p^{F}_{cond}$                & 0.3       &\cite{rocha2010information}  \\
		\cline{2-5}
		
	\end{tabular}
	\vspace{3mm}
	\caption{\textbf{Summary of the parameters of the model.} $H$ and $G$ stands for HIV and gonorrhea respectively, while $M$ and $F$ denote the gender dependency of the parameters.}
	\label{table:model parameters}
\end{table}

\noindent HIV dynamics is thus modeled as a Susceptible-Infected model for simplicity as it has been done in other analogous studies~\cite{macfadden_optimizing_2016,zhong_modeling_2018,tripathi2007modeling}, while gonorrhea dynamics is represented as a\\Susceptible~-~Exposed~-~Infected~-~Susceptible model. 

Transitions among the compartments are controlled by the transition probabilities listed in Table~\ref{table:model parameters}. 
At each time step (day), a susceptible $S$ individual having a sexual encounter can get infected by one or both gonorrhea and HIV if the partner is infectious. In particular, a susceptible $S$ individual can acquire gonorrhea and become $G_E$ or get infected by HIV and become $H$ with probability $\beta_G$ and $\beta_H$, respectively. Notice that there is also a joint probability of getting both diseases in the same time step and thus change state from $S$ to $HG_E$ (see Fig.~\ref{fig:model_diagram}). Unlike HIV, individuals that acquire gonorrhea become first exposed and then infectious. The exposed ones ($G_E$) have a probability $\epsilon_G$ of getting infectious which is inversely proportional to the average latency period~\cite{kretzschmar_modeling_1996}.
Individuals transit then from $G_E$ to the infectious state $G$ with a probability $\epsilon_G$.
We include in our model the fact that gonorrhea increases susceptibility to HIV~\cite{chesson_sexually_2000}, by multiplying the transition probability from $G_E$ to $HG$ and from $G$ to $HG$ by a factor $\theta$.

Finally, people with gonorrhea ($G$ and $HG$) can recover, becoming $S$ or $H$ again with probability $\gamma_{G}$. As before, this probability is inversely proportional to the respective average infectious period.

\medskip

In order to overcome the limited time span of the dataset, we impose periodic boundary conditions.
HIV spreads very slowly with respect to the daily time scale of the network, due to its low transmission probabilities. This assures that its spread will not critically depend on our specific choice of dataset extrapolation, as previous evidence shows~\cite{Stehle2011}.
To account for both FSWs and clients leaving and entering the system, we add vital dynamics.
More specifically, we assume that at each time step each individual has a certain probability of being replaced with a new one of the same type (FSWs with FSWs, and clients with clients). Moreover, each new individual enters the system with a probability of being infected equal to the initial prevalence of the diseases, to account for the fact that our network is not a closed system.
To fix the replacement probabilities, we wish to account for the fact that more active individuals might stay longer in the system. To this end, we divide individuals into three different activity classes, which are defined according to the number of sexual encounters: one encounter ($6 111$ individuals), between one and six ($6 443$), and more than six encounters ($2 229$).
For the second and third class, we calculate the activity time of each individual, i.e., the time between her/his first and last contact, focusing only on the individuals who had their first interaction in the first $200$ time steps. We assume that the replacement probability in each class is proportional to the inverse of its activity time, and we fix the proportionality constant so that we reach a steady endemic state for both diseases. This gives replacement probabilities of $4\cdot 10^{-4}$ and $1\cdot 10^{-4}$ for the second and third class, respectively. For the first class, for which we have no activity time, we fix $5\cdot 10^{-4}$, i.e., we assume that the individuals with the lowest activity gets replaced the most often.


\medskip

In each time step $t$, the total number of nodes $N$ is given by the sum of female (F) and male (M) individuals in all the compartments, which reads

\begin{equation}
\begin{aligned}
N=&\sum_{\alpha=F,M} S^{[\alpha]}(t)+G^{[\alpha]}_E(t)+G^{[\alpha]}(t)\\
&+H^{[\alpha]}(t)+HG^{[\alpha]}_E(t)+HG^{[\alpha]}(t) .
\end{aligned}
\end{equation}
Here, we define some macroscopic order parameters that we will use in the different scenarios. Specifically, we denote the density of HIV and gonorrhea infectious individuals of gender $\alpha$ as
\begin{equation}
h^{[\alpha]}(t)=\frac{H^{[\alpha]}(t)}{N^{[\alpha]}}
\end{equation}
\noindent and
\begin{equation}
g^{[\alpha]}(t)=\frac{G^{[\alpha]}(t)}{N^{[\alpha]}}.
\end{equation}

\noindent We denote with $h^{[\alpha]}$ and $g^{[\alpha]}$ their respective stationary state. Similarly, the global density of infectious individuals is given by

\begin{equation}\label{eq:rho}
\rho^{[\alpha]}(t)=h^{[\alpha]}(t)+g^{[\alpha]}(t)+hg^{[\alpha]}(t)
\end{equation}

\noindent with

\begin{equation}
hg^{[\alpha]}(t)=\frac{HG^{[\alpha]}(t)}{N^{[\alpha]}}, 
\end{equation}

\noindent being the density of individuals who are both gonorrhea-infectious and HIV-infectious. Again, we call $\rho^{[\alpha]}$ the stationary density of infectious for Eq.~\ref{eq:rho}.

\subsection*{Condom use, and risk compensation}
We assume the probability of condom use to be $p_{cond} = 0.3$ per sexual act. It corresponds to the average of what is measured in~\cite{rocha2010information}.
We interpret $p_{cond}$ as the ability of the FSW to enforce condom use during intercourse with her client. The lower its value, the lower her bargaining power.
We model the impact of risk compensation on condom use as a rescaling factor to $p_{cond}$. Therefore, in the absence of risk compensation ($0\%$), $p_{cond}$ remains unchanged when PrEP is used. On the other hand for $100\%$ we have full risk compensation and $p_{cond}$ goes to zero which means that FSWs on PrEP do not use condoms.

\subsection*{Modeling Setup}

In the analyses, we consider three scenarios, and the baseline. 
In the baseline scenario condoms are the only way of preventing HIV acquisition.
In all the other scenarios, we add PrEP with different sets of parameters. In the first scenario, we set high adoption ($50\%$ of women) and high PrEP efficacy ($70\%$ reduction in HIV susceptibility). In the second, we set low adoption ($10\%$ of women) keeping high PrEP efficacy ($70\%$ reduction in HIV susceptibility). In the third and last scenario, we set low adoption ($10\%$ of women) and low PrEP efficacy ($40\%$ reduction in HIV susceptibility). The different levels of PrEP efficacy are set to explore different scenarios of adherence~\cite{dimitrov2016prep}. We run these three scenarios for four different values of risk compensation: $0\%$ (no risk compensation), $33\%$, $66\%$, $100\%$ (full risk compensation).

We start by exploring the spread of HIV without gonorrhea in the first two scenarios. Each simulation starts with an initial prevalence of HIV equal to $10\%$, in both FSWs and clients. We run each scenario $20$ times, each over the course of $1.5\cdot 10^5$ days, to get sufficient statistics.
We then introduce gonorrhea and perform the simulations with the same number of runs and over the same time span.
In this case, as initial condition of each simulation, we set the prevalence of both HIV and gonorrhea equal to $10\%$, in both FSWs and clients.
We remark that initial disease prevalence are not the endemic prevalence of our system. They are only the initial conditions, and represent the baseline prevalence of the individuals entering the system. The final endemic prevalence are the result of the infection dynamics on the network.

\section*{Results}\label{results}

\begin{figure}[t]
	\centering
	\vspace{5mm}
	\includegraphics[width=.48\textwidth]{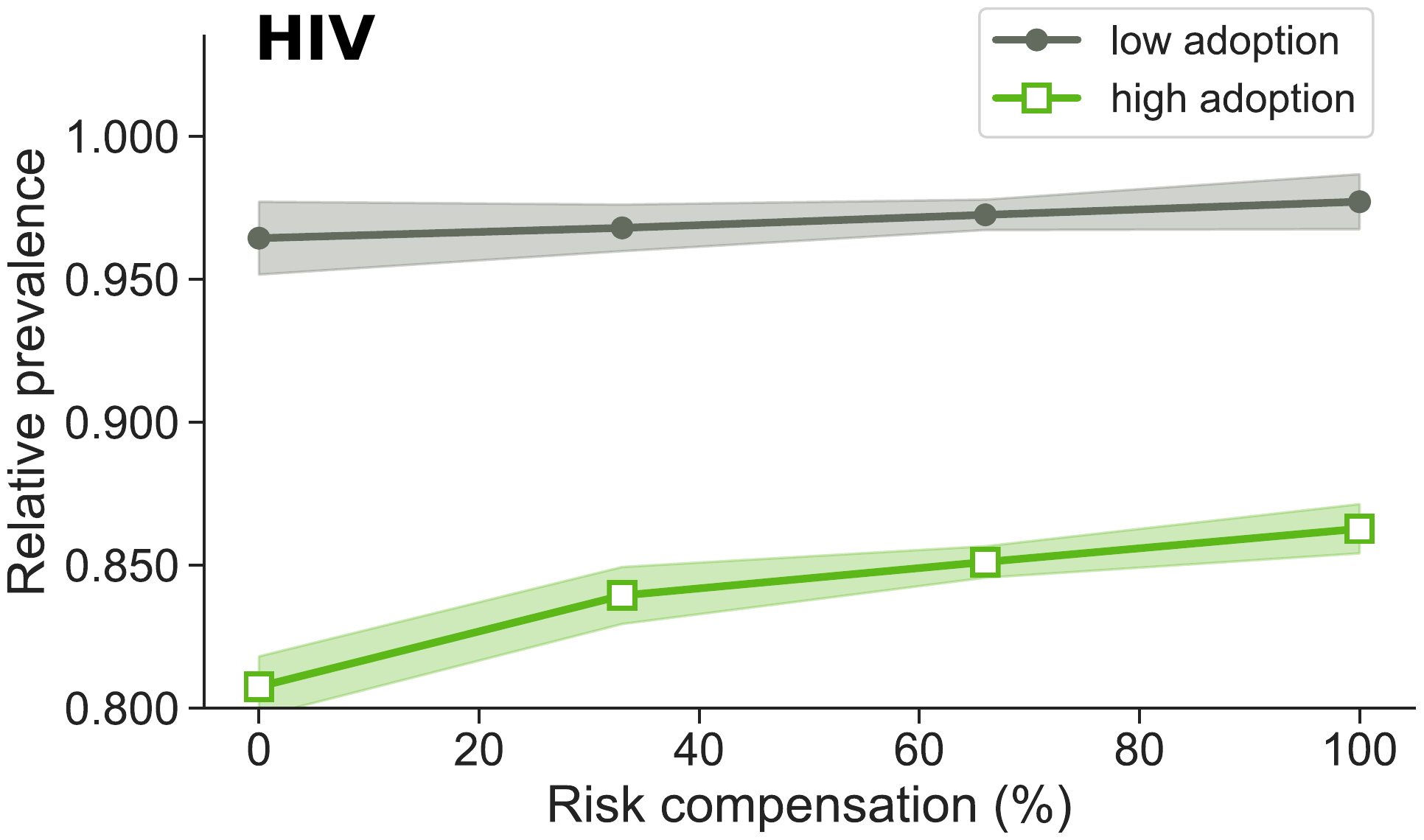}
	\caption{\textbf{HIV prevalence as a function of the risk compensation in the absence of gonorrhea for FSWs.} The black and green line indicate the scenarios of high ($50 \%$, scenario 1) and low adoption ($10 \%$, scenario 2) of sex workers use PrEP, with PrEP efficacy fixed to $70\%$. The solid line indicates the median value of the different realizations, while the area represents the standard deviation.}
	\label{fig:hiv_only}
\end{figure}

\subsection*{Uniform PrEP adoption}

In this section, we assign uniform probability of PrEP adoption among FSWs. 
For the first and second scenarios, without gonorrhea in the system, and for each value of risk compensation, we normalize the prevalence of HIV with the baseline in both FSWs and clients(see Fig.~\ref{fig:hiv_only}). 
As expected, we find that PrEP use decreases HIV prevalence in FSWs with respect to the baseline case. Notably, risk compensation has very little impact on HIV prevalence in both scenarios, as shown in Fig.~\ref{fig:hiv_only}.

\begin{figure*}[]
	\centering
	\includegraphics[width=0.99\textwidth]{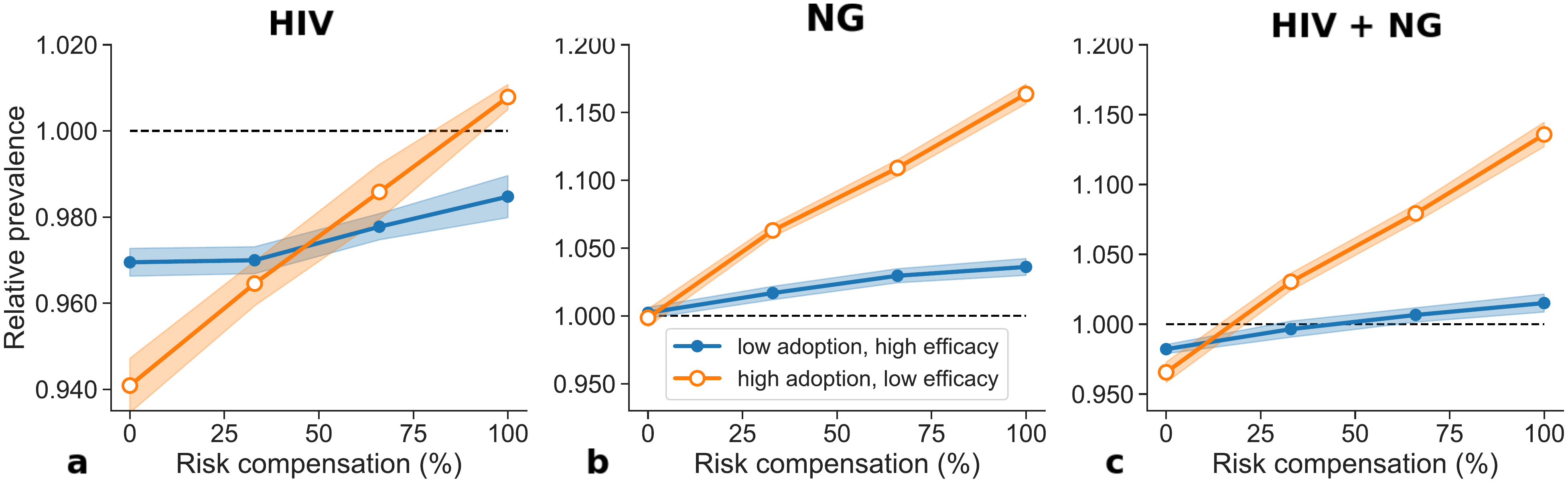}
	\vspace{3mm}
	\caption{\textbf{HIV and NG prevalence as a function of the risk compensation.} Relative prevalence of HIV (a), NG (b) and the combined incidence (c). The case of high and low adoption of PrEP corresponds to the ones of Fig.~\ref{fig:hiv_only}. Furthermore, we introduced two scenarios: one with high efficacy and one with low efficacy that lead to, respectively, $70 \%$ and $40 \%$ of reduction in the transmission probability. The solid lines indicate the median values of the different realizations, while the area represents the standard deviation. The dashed horizontal line denotes the baseline.}
	\label{fig:hiv_and_NG}
\end{figure*}

Next, we study the simultaneous spread of the two diseases. Figure~\ref{fig:hiv_and_NG} shows the impact on HIV and gonorrhea prevalence in FSWs, as well as the prevalence of co-infections.
We analyze the second (low PrEP adoption with high PrEP efficacy) and third (high PrEP adoption with low PrEP efficacy) scenarios.
With low adoption and high efficacy (blue lines), PrEP is able to constantly reduce HIV prevalence among FSWs, even in the case of full risk compensation.
The amount of such reduction is comparable to the case where gonorrhea is absent, and it is not affected by risk compensation, as seen by comparing Fig.~\ref{fig:hiv_only} and Fig.~\ref{fig:hiv_and_NG}.
This effect, driven by low adoption, is true in the cases of both high efficacy and low efficacy (not shown). Low PrEP efficacy simply decreases the overall impact of PrEP.
In addition, the scenario of low PrEP adoption has little effect on gonorrhea prevalence with respect to the baseline case.
However, we observe a slight decrease in co-infections (HIV + gonorrhea) when risk compensation is absent, or low.

We now turn to the scenario of high PrEP adoption with low PrEP efficacy (orange lines in Fig.~\ref{fig:hiv_and_NG}). In this case, both HIV prevalence and gonorrhea prevalence are sensitive to risk compensation. As expected, low risk compensation entails a drop in HIV prevalence, and has no negative effect on gonorrhea prevalence. On the other hand, it decreases co-infections, as in the other scenario (low PrEP adoption with high PrEP efficacy).
Increasing risk compensation sharply increases both HIV and gonorrhea prevalence, up to the case of full risk compensation, when HIV prevalence is actually slightly higher than the baseline. However, this simply represents the slightly lower efficacy of PrEP with respect to condom, hence it is sensitive to the specific value of PrEP efficacy set.
The effect of high risk compensation has instead a marked impact on gonorrhea, with the extreme case of full risk compensation showing a $15\%$ increase with respect to the baseline.

\subsection*{Targeted PrEP adoption}

\begin{figure*}[]
	\centering
	\includegraphics[width=0.99\textwidth]{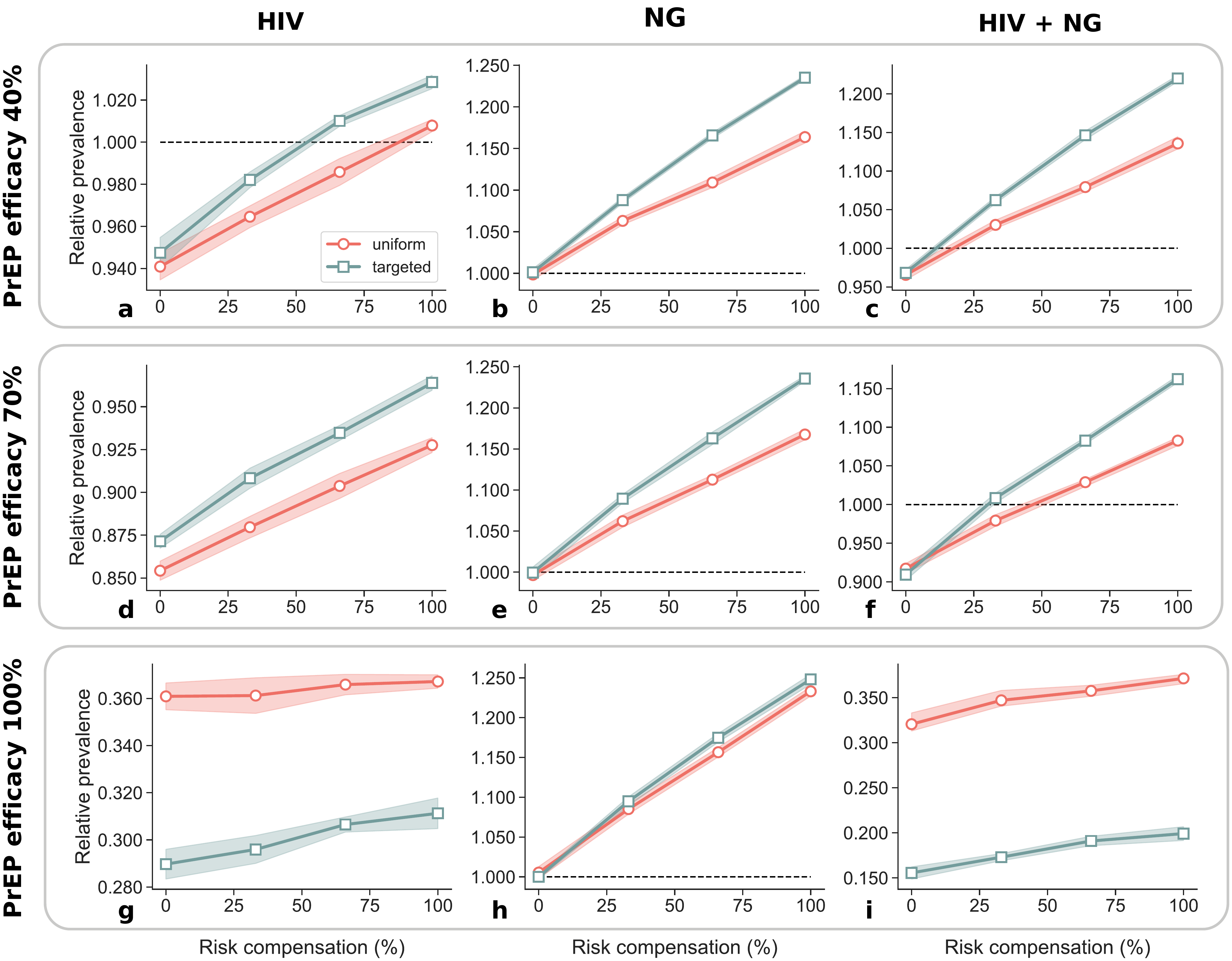}
	\vspace{3mm}
	\caption{\textbf{HIV and NG prevalence as a function of risk compensation as $50 \%$ of the sex workers are using PrEP.} Between the top (a-c), center (d-f) and bottom (g-i) panels the PrEP efficacy varies between $40 \%$, $70 \%$ and $100 \%$. The left (a,d,g), center (b,e,h) and right (c,f,i) columns show the HIV, NG and combined prevalence. The green and red lines indicate the prevalence as PrEP is distributed uniformly or in a targeted way, respectively. The solid line is the median prevalence over the different realizations, whereas the area indicates the variance. The dashed horizontal line denotes the baseline.}
	\label{fig:hiv_and_NG_interventions}
\end{figure*}

We now assume PrEP take-up happens preferentially among highly active FSWs, i.e., those with high numbers of sexual acts. Specifically, we divide FSWs in three classes: the $40\%$ least active, the $40\%$ mid active, and the $20\%$ most active. We then assume that women in the first class never adopt PrEP, women in the second with probability $0.81$, and women in the third one (the most active) always adopt it. These probabilities guarantee that the average adoption probability is $50\%$, to compare it with the uniform high PrEP adoption scenarios, i.e. scenarios 1 and 3.
In addition to the low PrEP efficacy ($40\%$, scenario 3), and high PrEP efficacy ($70\%$, scenario 1), we explore also perfect PrEP efficacy ($100\%$).
The results are shown in Fig.~\ref{fig:hiv_and_NG_interventions}.
The first column reports the impact on HIV prevalence. Clearly, we observe that increasing PrEP efficacy reduces HIV prevalence, for both uniform and targeted adoptions. Unexpectedly, for both efficacy $40\%$ (a) and $70\%$ (d) uniform PrEP adoption consistently outperforms targeted adoption, for every value of risk compensation. In particular, targeted intervention has a detrimental effect for high risk compensation, pushing HIV prevalence significantly above the baseline in the case of $40\%$ PrEP adoption.
When PrEP efficacy is perfect, the drop in HIV prevalence is marked (see Fig.~\ref{fig:hiv_and_NG_interventions}c)). However, the performance of the two PrEP adoption strategies changes completely (see Fig.~\ref{fig:hiv_and_NG_interventions}g). Here we observe that targeted adoption performs much better than uniform adoption. At the same time, risk compensation plays almost no role, as women on PrEP are protected no matter their condom use.

The effect of immunization strategies on gonorrhea prevalence is shown in Fig.~\ref{fig:hiv_and_NG_interventions}b,e,h. Gonorrhea prevalence is consistently higher than baseline for nonzero risk compensation. We note that the targeted intervention always increases gonorrhea prevalence more than uniform, albeit the difference between the two interventions decreases from lower to higher PrEP efficacy, becoming almost zero in case of perfect efficacy. This is due to the fact that targeted adoption is more sensitive to risk compensation than uniform adoption.

Finally, Fig.~\ref{fig:hiv_and_NG_interventions}c,f,i show the impact on co-infections. The overall behavior is driven by HIV prevalence (Fig.~\ref{fig:hiv_and_NG_interventions}a,b,c). However, for low risk compensation co-infections are consistently lower than baseline. 

\begin{figure}[]
	\centering
	\includegraphics[width=0.45\textwidth]{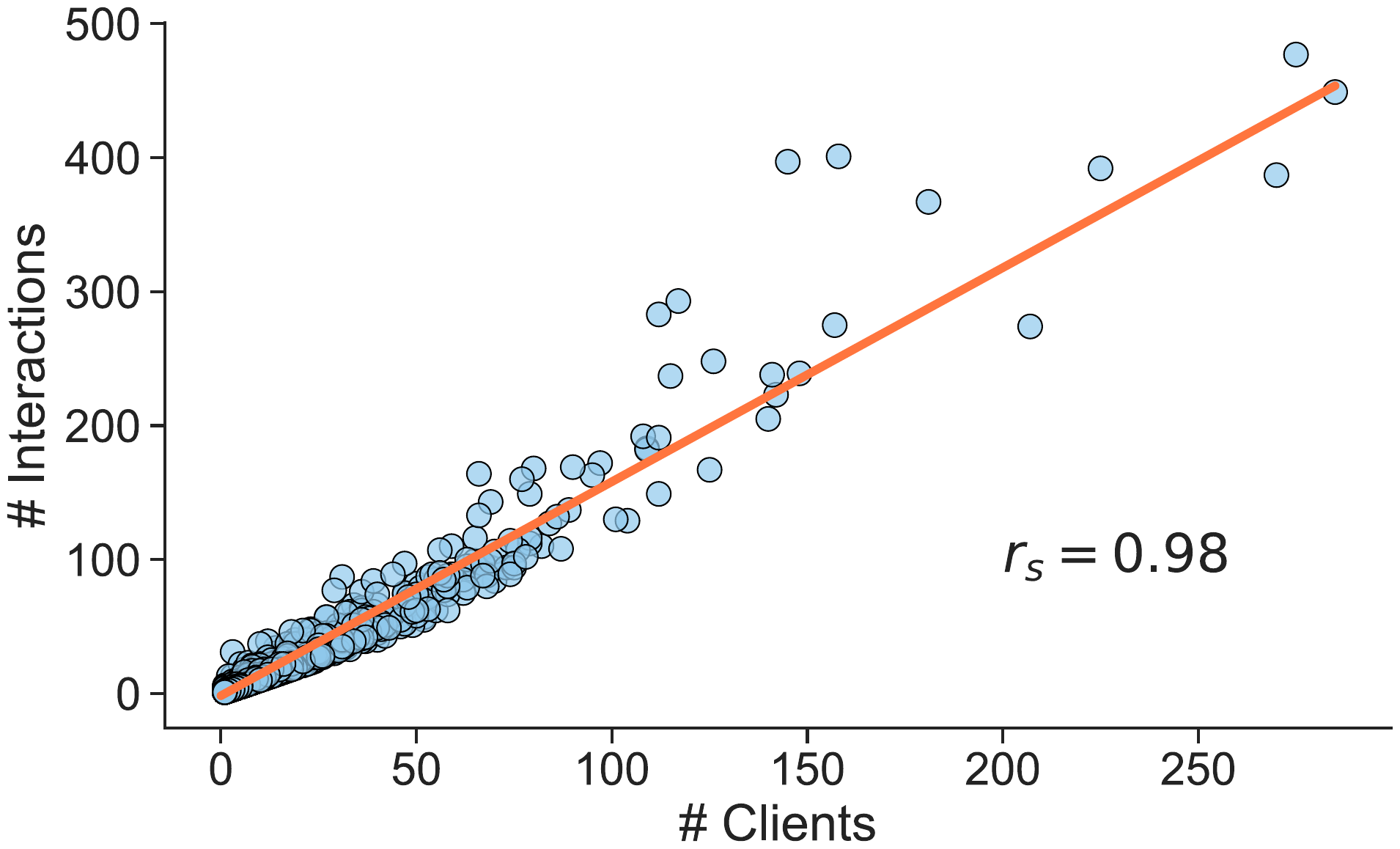}
	\vspace{3mm}
	\caption{\textbf{Number of interactions of sex workers as a function of their number of clients.} The solid red line indicates a linear fit of the data. The two are highly correlated, which is illustrated by the Spearman rank coefficient of $r_s = 0.98 \pm 0.0025$.}
	\label{fig:correlation}
\end{figure}

Instead of targeting the most active FSWs, another strategy would be to target the ones with the highest number of sexual partners. However, we find that the two quantities are extremely correlated, as Fig.~\ref{fig:correlation} shows, and we do not exchange any different behavior.

\medskip

\section*{Discussion}

We first considered circulation of HIV only, and examined the impact of PrEP adoption in a population of FSWs. We compared the prevalence of HIV with respect to the baseline scenario of condom use and no PrEP. We investigated the impact of risk compensation among FSWs, in terms of a potential reduction in condom use following PrEP adoption.
In this scenario, we found that risk compensation has a limited effect and does not seem to compromise the high efficacy of PrEP in reducing HIV prevalence.

We then studied co-circulation of HIV and gonorrhea, observing a complex interplay among the two diseases on one hand, and the two prophylactic tools on the other (condom, PrEP). Notably, we found two opposite regimes. In case of low risk compensation (women tend not to change their behavior towards condoms when they are on PrEP), our model suggests that providing PrEP to more women is efficient in reducing HIV prevalence.
Instead, in case of high risk compensation, HIV reduction requires interventions focusing on increasing PrEP efficacy in users. This can be attained with supportive structures and formative activities~\cite{reza2016prioritizing, eakle_i_2019, eakle2018designing}, aimed at increasing PrEP knowledge, and fighting logistical barriers and stigma~\cite{eakle_i_2019, ortblad2018acceptability}.
Our results however show that the existence of risk compensation increases gonorrhea prevalence, often even at low levels. This implies that PrEP diffusion strategies should entail consistent screenings to minimize the impact of this effect~\cite{jenness2017incidencegonochlamyinHIVpopMSM}.

Finally, we analysed the impact of non-uniform PrEP adoption strategies.
We compared the random distribution of PrEP with a targeted distribution to the most active FSWs. 
Surprisingly, and in contrast to previous findings~\cite{holme2017cost}, uniform distribution of PrEP proves more effective than the targeted distribution. This difference with previous studies can be explained by the fact that they only focused on perfect immunization (in our case translating into $100\%$ effective prophylaxis). Actually, in the case of imperfect immunization, there is the possibility of saturation effects. In other words, very active FSWs have such a high probability of eventually acquiring HIV, that an imperfect protection cannot provide substantial protection.
Accordingly, it proves more beneficial to target FSWs which are less active, but whose infection probability is substantially decreased by PrEP. Obviously, this argument does not apply if immunization is perfect, as our findings show.
Previous studies on immunization essentially searched for local rules, which serve as a proxy to identify highly connected nodes in the network~\cite{Cohen2003,holme2017cost}. However, as our results indicate, such protocols may be obsolete in the case of PrEP, since random immunization proves more efficient. 

Our work has several limitations. Firstly, we consider simplified compartmental models of both HIV and gonorrhea. We chose this as we were interested in performing scenario analyses rather than quantitative predictions, and so that we could rely on a minimal number of parameters. Secondly, the dataset on contacts among FSWs and their clients is not complete, as it does not contains non-commercial sexual acts, and might miss a certain number of non-reported commercial ones.
Finally, we do not include neither HIV nor gonorrhea testing. Consequently, we do not include the impact of treatment-as-prevention (TasP) as a measure to reduce HIV prevalence. While this seems unrealistic, we preferred to focus on prophylactic measures that directly prevent infections among FSWs, and study their already complex interactions. Notwithstanding, we believe that future scenario analyses should include testing and treatment in order to provide more realistic predictions.

\begin{acknowledgments}
	This work is the output of the Complexity72h workshop, held at IMT School in Lucca, Italy, 17-21 June
	2019 {\it https://complexity72h.weebly.com/}. 
	
	AST acknowledges FCT-Portugal for funding through projects PTDC/EEI-SII/1937/2014, UID/CEC/00408/2019 and UID/CEC/50021/2019.
	BS acknowledges financial support from the European Union's Horizon 2020 research and innovation programme under the Marie Skłodowska-Curie grant agreement No. 713679 and from the Universitat Rovira i Virgili (URV).
\end{acknowledgments}

%


\begin{thebibliography}{50}%
	\makeatletter
	\providecommand \@ifxundefined [1]{%
		\@ifx{#1\undefined}
	}%
	\providecommand \@ifnum [1]{%
		\ifnum #1\expandafter \@firstoftwo
		\else \expandafter \@secondoftwo
		\fi
	}%
	\providecommand \@ifx [1]{%
		\ifx #1\expandafter \@firstoftwo
		\else \expandafter \@secondoftwo
		\fi
	}%
	\providecommand \natexlab [1]{#1}%
	\providecommand \enquote  [1]{``#1''}%
	\providecommand \bibnamefont  [1]{#1}%
	\providecommand \bibfnamefont [1]{#1}%
	\providecommand \citenamefont [1]{#1}%
	\providecommand \href@noop [0]{\@secondoftwo}%
	\providecommand \href [0]{\begingroup \@sanitize@url \@href}%
	\providecommand \@href[1]{\@@startlink{#1}\@@href}%
	\providecommand \@@href[1]{\endgroup#1\@@endlink}%
	\providecommand \@sanitize@url [0]{\catcode `\\12\catcode `\$12\catcode
		`\&12\catcode `\#12\catcode `\^12\catcode `\_12\catcode `\%12\relax}%
	\providecommand \@@startlink[1]{}%
	\providecommand \@@endlink[0]{}%
	\providecommand \url  [0]{\begingroup\@sanitize@url \@url }%
	\providecommand \@url [1]{\endgroup\@href {#1}{\urlprefix }}%
	\providecommand \urlprefix  [0]{URL }%
	\providecommand \Eprint [0]{\href }%
	\providecommand \doibase [0]{http://dx.doi.org/}%
	\providecommand \selectlanguage [0]{\@gobble}%
	\providecommand \bibinfo  [0]{\@secondoftwo}%
	\providecommand \bibfield  [0]{\@secondoftwo}%
	\providecommand \translation [1]{[#1]}%
	\providecommand \BibitemOpen [0]{}%
	\providecommand \bibitemStop [0]{}%
	\providecommand \bibitemNoStop [0]{.\EOS\space}%
	\providecommand \EOS [0]{\spacefactor3000\relax}%
	\providecommand \BibitemShut  [1]{\csname bibitem#1\endcsname}%
	\let\auto@bib@innerbib\@empty
	\bibitem [{\citenamefont {Organization}\ \emph {et~al.}(2012)\citenamefont
		{Organization} \emph {et~al.}}]{world2012prevention}%
	\BibitemOpen
	\bibfield  {author} {\bibinfo {author} {\bibfnamefont {W.~H.}\ \bibnamefont
			{Organization}} \emph {et~al.},\ }\href@noop {} {\  (\bibinfo {year}
		{2012})}\BibitemShut {NoStop}%
	\bibitem [{\citenamefont {Baral}\ \emph {et~al.}(2012)\citenamefont {Baral},
		\citenamefont {Beyrer}, \citenamefont {Muessig}, \citenamefont {Poteat},
		\citenamefont {Wirtz}, \citenamefont {Decker}, \citenamefont {Sherman},\ and\
		\citenamefont {Kerrigan}}]{baral2012burden}%
	\BibitemOpen
	\bibfield  {author} {\bibinfo {author} {\bibfnamefont {S.}~\bibnamefont
			{Baral}}, \bibinfo {author} {\bibfnamefont {C.}~\bibnamefont {Beyrer}},
		\bibinfo {author} {\bibfnamefont {K.}~\bibnamefont {Muessig}}, \bibinfo
		{author} {\bibfnamefont {T.}~\bibnamefont {Poteat}}, \bibinfo {author}
		{\bibfnamefont {A.~L.}\ \bibnamefont {Wirtz}}, \bibinfo {author}
		{\bibfnamefont {M.~R.}\ \bibnamefont {Decker}}, \bibinfo {author}
		{\bibfnamefont {S.~G.}\ \bibnamefont {Sherman}}, \ and\ \bibinfo {author}
		{\bibfnamefont {D.}~\bibnamefont {Kerrigan}},\ }\href@noop {} {\bibfield
		{journal} {\bibinfo  {journal} {The Lancet infectious diseases}\ }\textbf
		{\bibinfo {volume} {12}},\ \bibinfo {pages} {538} (\bibinfo {year}
		{2012})}\BibitemShut {NoStop}%
	\bibitem [{\citenamefont {Shannon}\ \emph {et~al.}(2015)\citenamefont
		{Shannon}, \citenamefont {Strathdee}, \citenamefont {Goldenberg},
		\citenamefont {Duff}, \citenamefont {Mwangi}, \citenamefont {Rusakova},
		\citenamefont {Reza-Paul}, \citenamefont {Lau}, \citenamefont {Deering},
		\citenamefont {Pickles},\ and\ \citenamefont {Boily}}]{shannon_global_2015}%
	\BibitemOpen
	\bibfield  {author} {\bibinfo {author} {\bibfnamefont {K.}~\bibnamefont
			{Shannon}}, \bibinfo {author} {\bibfnamefont {S.~A.}\ \bibnamefont
			{Strathdee}}, \bibinfo {author} {\bibfnamefont {S.~M.}\ \bibnamefont
			{Goldenberg}}, \bibinfo {author} {\bibfnamefont {P.}~\bibnamefont {Duff}},
		\bibinfo {author} {\bibfnamefont {P.}~\bibnamefont {Mwangi}}, \bibinfo
		{author} {\bibfnamefont {M.}~\bibnamefont {Rusakova}}, \bibinfo {author}
		{\bibfnamefont {S.}~\bibnamefont {Reza-Paul}}, \bibinfo {author}
		{\bibfnamefont {J.}~\bibnamefont {Lau}}, \bibinfo {author} {\bibfnamefont
			{K.}~\bibnamefont {Deering}}, \bibinfo {author} {\bibfnamefont {M.~R.}\
			\bibnamefont {Pickles}}, \ and\ \bibinfo {author} {\bibfnamefont {M.-C.}\
			\bibnamefont {Boily}},\ }\href@noop {} {\bibfield  {journal} {\bibinfo
			{journal} {The Lancet}\ }\textbf {\bibinfo {volume} {385}},\ \bibinfo {pages}
		{55} (\bibinfo {year} {2015})}\BibitemShut {NoStop}%
	\bibitem [{\citenamefont {Ghimire}\ \emph {et~al.}(2011)\citenamefont
		{Ghimire}, \citenamefont {Smith}, \citenamefont {van Teijlingen},
		\citenamefont {Dahal},\ and\ \citenamefont {Luitel}}]{ghimire_reasons_2011}%
	\BibitemOpen
	\bibfield  {author} {\bibinfo {author} {\bibfnamefont {L.}~\bibnamefont
			{Ghimire}}, \bibinfo {author} {\bibfnamefont {W.~C.~S.}\ \bibnamefont
			{Smith}}, \bibinfo {author} {\bibfnamefont {E.~R.}\ \bibnamefont {van
				Teijlingen}}, \bibinfo {author} {\bibfnamefont {R.}~\bibnamefont {Dahal}}, \
		and\ \bibinfo {author} {\bibfnamefont {N.~P.}\ \bibnamefont {Luitel}},\
	}\href@noop {} {\bibfield  {journal} {\bibinfo  {journal} {BMC Women's
				Health}\ }\textbf {\bibinfo {volume} {11}},\ \bibinfo {pages} {42} (\bibinfo
		{year} {2011})}\BibitemShut {NoStop}%
	\bibitem [{\citenamefont {Shannon}\ \emph {et~al.}(2009)\citenamefont
		{Shannon}, \citenamefont {Strathdee}, \citenamefont {Shoveller},
		\citenamefont {Rusch}, \citenamefont {Kerr},\ and\ \citenamefont
		{Tyndall}}]{shannon2009structural}%
	\BibitemOpen
	\bibfield  {author} {\bibinfo {author} {\bibfnamefont {K.}~\bibnamefont
			{Shannon}}, \bibinfo {author} {\bibfnamefont {S.~A.}\ \bibnamefont
			{Strathdee}}, \bibinfo {author} {\bibfnamefont {J.}~\bibnamefont
			{Shoveller}}, \bibinfo {author} {\bibfnamefont {M.}~\bibnamefont {Rusch}},
		\bibinfo {author} {\bibfnamefont {T.}~\bibnamefont {Kerr}}, \ and\ \bibinfo
		{author} {\bibfnamefont {M.~W.}\ \bibnamefont {Tyndall}},\ }\href@noop {}
	{\bibfield  {journal} {\bibinfo  {journal} {American journal of public
				health}\ }\textbf {\bibinfo {volume} {99}},\ \bibinfo {pages} {659} (\bibinfo
		{year} {2009})}\BibitemShut {NoStop}%
	\bibitem [{\citenamefont {Urada}\ \emph {et~al.}(2012)\citenamefont {Urada},
		\citenamefont {Morisky}, \citenamefont {Pimentel-Simbulan}, \citenamefont
		{Silverman},\ and\ \citenamefont {Strathdee}}]{urada2012condom}%
	\BibitemOpen
	\bibfield  {author} {\bibinfo {author} {\bibfnamefont {L.~A.}\ \bibnamefont
			{Urada}}, \bibinfo {author} {\bibfnamefont {D.~E.}\ \bibnamefont {Morisky}},
		\bibinfo {author} {\bibfnamefont {N.}~\bibnamefont {Pimentel-Simbulan}},
		\bibinfo {author} {\bibfnamefont {J.~G.}\ \bibnamefont {Silverman}}, \ and\
		\bibinfo {author} {\bibfnamefont {S.~A.}\ \bibnamefont {Strathdee}},\
	}\href@noop {} {\bibfield  {journal} {\bibinfo  {journal} {PLoS One}\
		}\textbf {\bibinfo {volume} {7}},\ \bibinfo {pages} {e33282} (\bibinfo {year}
		{2012})}\BibitemShut {NoStop}%
	\bibitem [{\citenamefont {Eakle}\ \emph {et~al.}(2019)\citenamefont {Eakle},
		\citenamefont {Bothma}, \citenamefont {Bourne}, \citenamefont {Gumede},
		\citenamefont {Motsosi},\ and\ \citenamefont {Rees}}]{eakle_i_2019}%
	\BibitemOpen
	\bibfield  {author} {\bibinfo {author} {\bibfnamefont {R.}~\bibnamefont
			{Eakle}}, \bibinfo {author} {\bibfnamefont {R.}~\bibnamefont {Bothma}},
		\bibinfo {author} {\bibfnamefont {A.}~\bibnamefont {Bourne}}, \bibinfo
		{author} {\bibfnamefont {S.}~\bibnamefont {Gumede}}, \bibinfo {author}
		{\bibfnamefont {K.}~\bibnamefont {Motsosi}}, \ and\ \bibinfo {author}
		{\bibfnamefont {H.}~\bibnamefont {Rees}},\ }\href@noop {} {\bibfield
		{journal} {\bibinfo  {journal} {PLOS ONE}\ }\textbf {\bibinfo {volume}
			{14}},\ \bibinfo {pages} {e0212271} (\bibinfo {year} {2019})}\BibitemShut
	{NoStop}%
	\bibitem [{\citenamefont {Flash}\ \emph {et~al.}(2017)\citenamefont {Flash},
		\citenamefont {Dale},\ and\ \citenamefont
		{Krakower}}]{flash_pre-exposure_2017}%
	\BibitemOpen
	\bibfield  {author} {\bibinfo {author} {\bibfnamefont {C.~A.}\ \bibnamefont
			{Flash}}, \bibinfo {author} {\bibfnamefont {S.~K.}\ \bibnamefont {Dale}}, \
		and\ \bibinfo {author} {\bibfnamefont {D.~S.}\ \bibnamefont {Krakower}},\
	}\href@noop {} {\bibfield  {journal} {\bibinfo  {journal} {International
				Journal of Women's Health}\ }\textbf {\bibinfo {volume} {9}},\ \bibinfo
		{pages} {391} (\bibinfo {year} {2017})}\BibitemShut {NoStop}%
	\bibitem [{\citenamefont {Fonner}\ \emph {et~al.}(2016)\citenamefont {Fonner},
		\citenamefont {Dalglish}, \citenamefont {Kennedy}, \citenamefont {Baggaley},
		\citenamefont {O’reilly}, \citenamefont {Koechlin}, \citenamefont
		{Rodolph}, \citenamefont {Hodges-Mameletzis},\ and\ \citenamefont
		{Grant}}]{fonner2016effectiveness}%
	\BibitemOpen
	\bibfield  {author} {\bibinfo {author} {\bibfnamefont {V.~A.}\ \bibnamefont
			{Fonner}}, \bibinfo {author} {\bibfnamefont {S.~L.}\ \bibnamefont
			{Dalglish}}, \bibinfo {author} {\bibfnamefont {C.~E.}\ \bibnamefont
			{Kennedy}}, \bibinfo {author} {\bibfnamefont {R.}~\bibnamefont {Baggaley}},
		\bibinfo {author} {\bibfnamefont {K.~R.}\ \bibnamefont {O’reilly}},
		\bibinfo {author} {\bibfnamefont {F.~M.}\ \bibnamefont {Koechlin}}, \bibinfo
		{author} {\bibfnamefont {M.}~\bibnamefont {Rodolph}}, \bibinfo {author}
		{\bibfnamefont {I.}~\bibnamefont {Hodges-Mameletzis}}, \ and\ \bibinfo
		{author} {\bibfnamefont {R.~M.}\ \bibnamefont {Grant}},\ }\href@noop {}
	{\bibfield  {journal} {\bibinfo  {journal} {AIDS (London, England)}\ }\textbf
		{\bibinfo {volume} {30}},\ \bibinfo {pages} {1973} (\bibinfo {year}
		{2016})}\BibitemShut {NoStop}%
	\bibitem [{\citenamefont {Galea}\ \emph {et~al.}(2011)\citenamefont {Galea},
		\citenamefont {Kinsler}, \citenamefont {Salazar}, \citenamefont {Lee},
		\citenamefont {Giron}, \citenamefont {Sayles}, \citenamefont {C{\'a}ceres},\
		and\ \citenamefont {Cunningham}}]{galea2011acceptability}%
	\BibitemOpen
	\bibfield  {author} {\bibinfo {author} {\bibfnamefont {J.~T.}\ \bibnamefont
			{Galea}}, \bibinfo {author} {\bibfnamefont {J.~J.}\ \bibnamefont {Kinsler}},
		\bibinfo {author} {\bibfnamefont {X.}~\bibnamefont {Salazar}}, \bibinfo
		{author} {\bibfnamefont {S.-J.}\ \bibnamefont {Lee}}, \bibinfo {author}
		{\bibfnamefont {M.}~\bibnamefont {Giron}}, \bibinfo {author} {\bibfnamefont
			{J.~N.}\ \bibnamefont {Sayles}}, \bibinfo {author} {\bibfnamefont
			{C.}~\bibnamefont {C{\'a}ceres}}, \ and\ \bibinfo {author} {\bibfnamefont
			{W.~E.}\ \bibnamefont {Cunningham}},\ }\href@noop {} {\bibfield  {journal}
		{\bibinfo  {journal} {International journal of STD \& AIDS}\ }\textbf
		{\bibinfo {volume} {22}},\ \bibinfo {pages} {256} (\bibinfo {year}
		{2011})}\BibitemShut {NoStop}%
	\bibitem [{\citenamefont {Reza-Paul}\ \emph {et~al.}(2016)\citenamefont
		{Reza-Paul}, \citenamefont {Lazarus}, \citenamefont {Doshi}, \citenamefont
		{Rahman}, \citenamefont {Ramaiah}, \citenamefont {Maiya}, \citenamefont
		{Venugopal}, \citenamefont {Venukumar}, \citenamefont {Sundararaman},
		\citenamefont {Becker} \emph {et~al.}}]{reza2016prioritizing}%
	\BibitemOpen
	\bibfield  {author} {\bibinfo {author} {\bibfnamefont {S.}~\bibnamefont
			{Reza-Paul}}, \bibinfo {author} {\bibfnamefont {L.}~\bibnamefont {Lazarus}},
		\bibinfo {author} {\bibfnamefont {M.}~\bibnamefont {Doshi}}, \bibinfo
		{author} {\bibfnamefont {S.~H.~U.}\ \bibnamefont {Rahman}}, \bibinfo {author}
		{\bibfnamefont {M.}~\bibnamefont {Ramaiah}}, \bibinfo {author} {\bibfnamefont
			{R.}~\bibnamefont {Maiya}}, \bibinfo {author} {\bibfnamefont
			{M.}~\bibnamefont {Venugopal}}, \bibinfo {author} {\bibfnamefont
			{K.}~\bibnamefont {Venukumar}}, \bibinfo {author} {\bibfnamefont
			{S.}~\bibnamefont {Sundararaman}}, \bibinfo {author} {\bibfnamefont
			{M.}~\bibnamefont {Becker}},  \emph {et~al.},\ }\href@noop {} {\bibfield
		{journal} {\bibinfo  {journal} {PloS one}\ }\textbf {\bibinfo {volume}
			{11}},\ \bibinfo {pages} {e0166889} (\bibinfo {year} {2016})}\BibitemShut
	{NoStop}%
	\bibitem [{\citenamefont {Eakle}\ \emph {et~al.}(2018)\citenamefont {Eakle},
		\citenamefont {Mutanha}, \citenamefont {Mbogua}, \citenamefont {Sibanyoni},
		\citenamefont {Bourne}, \citenamefont {Gomez}, \citenamefont {Venter},\ and\
		\citenamefont {Rees}}]{eakle2018designing}%
	\BibitemOpen
	\bibfield  {author} {\bibinfo {author} {\bibfnamefont {R.}~\bibnamefont
			{Eakle}}, \bibinfo {author} {\bibfnamefont {N.}~\bibnamefont {Mutanha}},
		\bibinfo {author} {\bibfnamefont {J.}~\bibnamefont {Mbogua}}, \bibinfo
		{author} {\bibfnamefont {M.}~\bibnamefont {Sibanyoni}}, \bibinfo {author}
		{\bibfnamefont {A.}~\bibnamefont {Bourne}}, \bibinfo {author} {\bibfnamefont
			{G.}~\bibnamefont {Gomez}}, \bibinfo {author} {\bibfnamefont
			{F.}~\bibnamefont {Venter}}, \ and\ \bibinfo {author} {\bibfnamefont
			{H.}~\bibnamefont {Rees}},\ }\href@noop {} {\bibfield  {journal} {\bibinfo
			{journal} {BMJ open}\ }\textbf {\bibinfo {volume} {8}},\ \bibinfo {pages}
		{e019292} (\bibinfo {year} {2018})}\BibitemShut {NoStop}%
	\bibitem [{\citenamefont {Ortblad}\ \emph {et~al.}(2018)\citenamefont
		{Ortblad}, \citenamefont {Chanda}, \citenamefont {Musoke}, \citenamefont
		{Ngabirano}, \citenamefont {Mwale}, \citenamefont {Nakitende}, \citenamefont
		{Chongo}, \citenamefont {Kamungoma}, \citenamefont {Kanchele}, \citenamefont
		{B{\"a}rnighausen} \emph {et~al.}}]{ortblad2018acceptability}%
	\BibitemOpen
	\bibfield  {author} {\bibinfo {author} {\bibfnamefont {K.~F.}\ \bibnamefont
			{Ortblad}}, \bibinfo {author} {\bibfnamefont {M.~M.}\ \bibnamefont {Chanda}},
		\bibinfo {author} {\bibfnamefont {D.~K.}\ \bibnamefont {Musoke}}, \bibinfo
		{author} {\bibfnamefont {T.}~\bibnamefont {Ngabirano}}, \bibinfo {author}
		{\bibfnamefont {M.}~\bibnamefont {Mwale}}, \bibinfo {author} {\bibfnamefont
			{A.}~\bibnamefont {Nakitende}}, \bibinfo {author} {\bibfnamefont
			{S.}~\bibnamefont {Chongo}}, \bibinfo {author} {\bibfnamefont
			{N.}~\bibnamefont {Kamungoma}}, \bibinfo {author} {\bibfnamefont
			{C.}~\bibnamefont {Kanchele}}, \bibinfo {author} {\bibfnamefont
			{T.}~\bibnamefont {B{\"a}rnighausen}},  \emph {et~al.},\ }\href@noop {}
	{\bibfield  {journal} {\bibinfo  {journal} {BMC infectious diseases}\
		}\textbf {\bibinfo {volume} {18}},\ \bibinfo {pages} {503} (\bibinfo {year}
		{2018})}\BibitemShut {NoStop}%
	\bibitem [{\citenamefont {Mboup}\ \emph {et~al.}(2018)\citenamefont {Mboup},
		\citenamefont {B{\'e}hanzin}, \citenamefont {Gu{\'e}dou}, \citenamefont
		{Geraldo}, \citenamefont {Goma-Mats{\'e}ts{\'e}}, \citenamefont
		{Gigu{\`e}re}, \citenamefont {Aza-Gnandji}, \citenamefont {Kessou},
		\citenamefont {Diallo}, \citenamefont {K{\^e}k{\^e}} \emph
		{et~al.}}]{mboup2018early}%
	\BibitemOpen
	\bibfield  {author} {\bibinfo {author} {\bibfnamefont {A.}~\bibnamefont
			{Mboup}}, \bibinfo {author} {\bibfnamefont {L.}~\bibnamefont {B{\'e}hanzin}},
		\bibinfo {author} {\bibfnamefont {F.~A.}\ \bibnamefont {Gu{\'e}dou}},
		\bibinfo {author} {\bibfnamefont {N.}~\bibnamefont {Geraldo}}, \bibinfo
		{author} {\bibfnamefont {E.}~\bibnamefont {Goma-Mats{\'e}ts{\'e}}}, \bibinfo
		{author} {\bibfnamefont {K.}~\bibnamefont {Gigu{\`e}re}}, \bibinfo {author}
		{\bibfnamefont {M.}~\bibnamefont {Aza-Gnandji}}, \bibinfo {author}
		{\bibfnamefont {L.}~\bibnamefont {Kessou}}, \bibinfo {author} {\bibfnamefont
			{M.}~\bibnamefont {Diallo}}, \bibinfo {author} {\bibfnamefont {R.~K.}\
			\bibnamefont {K{\^e}k{\^e}}},  \emph {et~al.},\ }\href@noop {} {\bibfield
		{journal} {\bibinfo  {journal} {Journal of the International AIDS Society}\
		}\textbf {\bibinfo {volume} {21}},\ \bibinfo {pages} {e25208} (\bibinfo
		{year} {2018})}\BibitemShut {NoStop}%
	\bibitem [{\citenamefont {Bazzi}\ \emph {et~al.}(2019)\citenamefont {Bazzi},
		\citenamefont {Yotebieng}, \citenamefont {Otticha}, \citenamefont {Rota},
		\citenamefont {Agot}, \citenamefont {Ohaga},\ and\ \citenamefont
		{Syvertsen}}]{bazzi2019pr}%
	\BibitemOpen
	\bibfield  {author} {\bibinfo {author} {\bibfnamefont {A.~R.}\ \bibnamefont
			{Bazzi}}, \bibinfo {author} {\bibfnamefont {K.}~\bibnamefont {Yotebieng}},
		\bibinfo {author} {\bibfnamefont {S.}~\bibnamefont {Otticha}}, \bibinfo
		{author} {\bibfnamefont {G.}~\bibnamefont {Rota}}, \bibinfo {author}
		{\bibfnamefont {K.}~\bibnamefont {Agot}}, \bibinfo {author} {\bibfnamefont
			{S.}~\bibnamefont {Ohaga}}, \ and\ \bibinfo {author} {\bibfnamefont {J.~L.}\
			\bibnamefont {Syvertsen}},\ }\href@noop {} {\bibfield  {journal} {\bibinfo
			{journal} {Journal of the international AIDS society}\ }\textbf {\bibinfo
			{volume} {22}},\ \bibinfo {pages} {e25266} (\bibinfo {year}
		{2019})}\BibitemShut {NoStop}%
	\bibitem [{\citenamefont {Grant}\ \emph {et~al.}(2017)\citenamefont {Grant},
		\citenamefont {Mukandavire}, \citenamefont {Eakle}, \citenamefont {Prudden},
		\citenamefont {B~Gomez}, \citenamefont {Rees},\ and\ \citenamefont
		{Watts}}]{grant_when_2017}%
	\BibitemOpen
	\bibfield  {author} {\bibinfo {author} {\bibfnamefont {H.}~\bibnamefont
			{Grant}}, \bibinfo {author} {\bibfnamefont {Z.}~\bibnamefont {Mukandavire}},
		\bibinfo {author} {\bibfnamefont {R.}~\bibnamefont {Eakle}}, \bibinfo
		{author} {\bibfnamefont {H.}~\bibnamefont {Prudden}}, \bibinfo {author}
		{\bibfnamefont {G.}~\bibnamefont {B~Gomez}}, \bibinfo {author} {\bibfnamefont
			{H.}~\bibnamefont {Rees}}, \ and\ \bibinfo {author} {\bibfnamefont
			{C.}~\bibnamefont {Watts}},\ }\href@noop {} {\bibfield  {journal} {\bibinfo
			{journal} {Journal of the International AIDS Society}\ }\textbf {\bibinfo
			{volume} {20}} (\bibinfo {year} {2017})}\BibitemShut {NoStop}%
	\bibitem [{\citenamefont {Cassell}\ \emph {et~al.}(2006)\citenamefont
		{Cassell}, \citenamefont {Halperin}, \citenamefont {Shelton},\ and\
		\citenamefont {Stanton}}]{cassell2006risk}%
	\BibitemOpen
	\bibfield  {author} {\bibinfo {author} {\bibfnamefont {M.~M.}\ \bibnamefont
			{Cassell}}, \bibinfo {author} {\bibfnamefont {D.~T.}\ \bibnamefont
			{Halperin}}, \bibinfo {author} {\bibfnamefont {J.~D.}\ \bibnamefont
			{Shelton}}, \ and\ \bibinfo {author} {\bibfnamefont {D.}~\bibnamefont
			{Stanton}},\ }\href@noop {} {\bibfield  {journal} {\bibinfo  {journal} {Bmj}\
		}\textbf {\bibinfo {volume} {332}},\ \bibinfo {pages} {605} (\bibinfo {year}
		{2006})}\BibitemShut {NoStop}%
	\bibitem [{\citenamefont {Blumenthal}\ and\ \citenamefont
		{Haubrich}(2014)}]{blumenthal2014risk}%
	\BibitemOpen
	\bibfield  {author} {\bibinfo {author} {\bibfnamefont {J.}~\bibnamefont
			{Blumenthal}}\ and\ \bibinfo {author} {\bibfnamefont {R.}~\bibnamefont
			{Haubrich}},\ }\href@noop {} {\bibfield  {journal} {\bibinfo  {journal} {The
				virtual mentor: VM}\ }\textbf {\bibinfo {volume} {16}},\ \bibinfo {pages}
		{909} (\bibinfo {year} {2014})}\BibitemShut {NoStop}%
	\bibitem [{\citenamefont {Nguyen}\ \emph {et~al.}(2018)\citenamefont {Nguyen},
		\citenamefont {Greenwald}, \citenamefont {Trottier}, \citenamefont {Cadieux},
		\citenamefont {Goyette}, \citenamefont {Beauchemin}, \citenamefont {Charest},
		\citenamefont {Longpr{\'e}}, \citenamefont {Lavoie}, \citenamefont {Tossa}
		\emph {et~al.}}]{nguyen2018incidenceSTIbeforeafterprep}%
	\BibitemOpen
	\bibfield  {author} {\bibinfo {author} {\bibfnamefont {V.-K.}\ \bibnamefont
			{Nguyen}}, \bibinfo {author} {\bibfnamefont {Z.~R.}\ \bibnamefont
			{Greenwald}}, \bibinfo {author} {\bibfnamefont {H.}~\bibnamefont {Trottier}},
		\bibinfo {author} {\bibfnamefont {M.}~\bibnamefont {Cadieux}}, \bibinfo
		{author} {\bibfnamefont {A.}~\bibnamefont {Goyette}}, \bibinfo {author}
		{\bibfnamefont {M.}~\bibnamefont {Beauchemin}}, \bibinfo {author}
		{\bibfnamefont {L.}~\bibnamefont {Charest}}, \bibinfo {author} {\bibfnamefont
			{D.}~\bibnamefont {Longpr{\'e}}}, \bibinfo {author} {\bibfnamefont
			{S.}~\bibnamefont {Lavoie}}, \bibinfo {author} {\bibfnamefont {H.~G.}\
			\bibnamefont {Tossa}},  \emph {et~al.},\ }\href@noop {} {\bibfield  {journal}
		{\bibinfo  {journal} {AIDS (London, England)}\ }\textbf {\bibinfo {volume}
			{32}},\ \bibinfo {pages} {523} (\bibinfo {year} {2018})}\BibitemShut
	{NoStop}%
	\bibitem [{\citenamefont {Kalichman}\ \emph {et~al.}(2011)\citenamefont
		{Kalichman}, \citenamefont {Pellowski},\ and\ \citenamefont
		{Turner}}]{kalichman2011prevalenceSTIcoinfinHIVpos}%
	\BibitemOpen
	\bibfield  {author} {\bibinfo {author} {\bibfnamefont {S.~C.}\ \bibnamefont
			{Kalichman}}, \bibinfo {author} {\bibfnamefont {J.}~\bibnamefont
			{Pellowski}}, \ and\ \bibinfo {author} {\bibfnamefont {C.}~\bibnamefont
			{Turner}},\ }\href@noop {} {\bibfield  {journal} {\bibinfo  {journal}
			{Sexually transmitted infections}\ }\textbf {\bibinfo {volume} {87}},\
		\bibinfo {pages} {183} (\bibinfo {year} {2011})}\BibitemShut {NoStop}%
	\bibitem [{\citenamefont {Mugwanya}\ \emph {et~al.}(2013)\citenamefont
		{Mugwanya}, \citenamefont {Donnell}, \citenamefont {Celum}, \citenamefont
		{Thomas}, \citenamefont {Ndase}, \citenamefont {Mugo}, \citenamefont
		{Katabira}, \citenamefont {Ngure}, \citenamefont {Baeten}, \citenamefont
		{Team} \emph {et~al.}}]{mugwanya2013sexual}%
	\BibitemOpen
	\bibfield  {author} {\bibinfo {author} {\bibfnamefont {K.~K.}\ \bibnamefont
			{Mugwanya}}, \bibinfo {author} {\bibfnamefont {D.}~\bibnamefont {Donnell}},
		\bibinfo {author} {\bibfnamefont {C.}~\bibnamefont {Celum}}, \bibinfo
		{author} {\bibfnamefont {K.~K.}\ \bibnamefont {Thomas}}, \bibinfo {author}
		{\bibfnamefont {P.}~\bibnamefont {Ndase}}, \bibinfo {author} {\bibfnamefont
			{N.}~\bibnamefont {Mugo}}, \bibinfo {author} {\bibfnamefont {E.}~\bibnamefont
			{Katabira}}, \bibinfo {author} {\bibfnamefont {K.}~\bibnamefont {Ngure}},
		\bibinfo {author} {\bibfnamefont {J.~M.}\ \bibnamefont {Baeten}}, \bibinfo
		{author} {\bibfnamefont {P.~P.~S.}\ \bibnamefont {Team}},  \emph {et~al.},\
	}\href@noop {} {\bibfield  {journal} {\bibinfo  {journal} {The Lancet
				infectious diseases}\ }\textbf {\bibinfo {volume} {13}},\ \bibinfo {pages}
		{1021} (\bibinfo {year} {2013})}\BibitemShut {NoStop}%
	\bibitem [{\citenamefont {Pilgrim}\ \emph {et~al.}(2018)\citenamefont
		{Pilgrim}, \citenamefont {Jani}, \citenamefont {Mathur}, \citenamefont
		{Kahabuka}, \citenamefont {Saria}, \citenamefont {Makyao}, \citenamefont
		{Apicella},\ and\ \citenamefont {Pulerwitz}}]{pilgrim2018provider}%
	\BibitemOpen
	\bibfield  {author} {\bibinfo {author} {\bibfnamefont {N.}~\bibnamefont
			{Pilgrim}}, \bibinfo {author} {\bibfnamefont {N.}~\bibnamefont {Jani}},
		\bibinfo {author} {\bibfnamefont {S.}~\bibnamefont {Mathur}}, \bibinfo
		{author} {\bibfnamefont {C.}~\bibnamefont {Kahabuka}}, \bibinfo {author}
		{\bibfnamefont {V.}~\bibnamefont {Saria}}, \bibinfo {author} {\bibfnamefont
			{N.}~\bibnamefont {Makyao}}, \bibinfo {author} {\bibfnamefont
			{L.}~\bibnamefont {Apicella}}, \ and\ \bibinfo {author} {\bibfnamefont
			{J.}~\bibnamefont {Pulerwitz}},\ }\href@noop {} {\bibfield  {journal}
		{\bibinfo  {journal} {PloS one}\ }\textbf {\bibinfo {volume} {13}},\ \bibinfo
		{pages} {e0196280} (\bibinfo {year} {2018})}\BibitemShut {NoStop}%
	\bibitem [{\citenamefont {Unemo}\ and\ \citenamefont
		{Shafer}(2014)}]{unemo2014AMRgono}%
	\BibitemOpen
	\bibfield  {author} {\bibinfo {author} {\bibfnamefont {M.}~\bibnamefont
			{Unemo}}\ and\ \bibinfo {author} {\bibfnamefont {W.~M.}\ \bibnamefont
			{Shafer}},\ }\href@noop {} {\bibfield  {journal} {\bibinfo  {journal}
			{Clinical microbiology reviews}\ }\textbf {\bibinfo {volume} {27}},\ \bibinfo
		{pages} {587} (\bibinfo {year} {2014})}\BibitemShut {NoStop}%
	\bibitem [{\citenamefont {Jenness}\ \emph {et~al.}(2017)\citenamefont
		{Jenness}, \citenamefont {Weiss}, \citenamefont {Goodreau}, \citenamefont
		{Gift}, \citenamefont {Chesson}, \citenamefont {Hoover}, \citenamefont
		{Smith}, \citenamefont {Liu}, \citenamefont {Sullivan},\ and\ \citenamefont
		{Rosenberg}}]{jenness2017incidencegonochlamyinHIVpopMSM}%
	\BibitemOpen
	\bibfield  {author} {\bibinfo {author} {\bibfnamefont {S.~M.}\ \bibnamefont
			{Jenness}}, \bibinfo {author} {\bibfnamefont {K.~M.}\ \bibnamefont {Weiss}},
		\bibinfo {author} {\bibfnamefont {S.~M.}\ \bibnamefont {Goodreau}}, \bibinfo
		{author} {\bibfnamefont {T.}~\bibnamefont {Gift}}, \bibinfo {author}
		{\bibfnamefont {H.}~\bibnamefont {Chesson}}, \bibinfo {author} {\bibfnamefont
			{K.~W.}\ \bibnamefont {Hoover}}, \bibinfo {author} {\bibfnamefont {D.~K.}\
			\bibnamefont {Smith}}, \bibinfo {author} {\bibfnamefont {A.~Y.}\ \bibnamefont
			{Liu}}, \bibinfo {author} {\bibfnamefont {P.~S.}\ \bibnamefont {Sullivan}}, \
		and\ \bibinfo {author} {\bibfnamefont {E.~S.}\ \bibnamefont {Rosenberg}},\
	}\href@noop {} {\bibfield  {journal} {\bibinfo  {journal} {Clinical
				Infectious Diseases}\ }\textbf {\bibinfo {volume} {65}},\ \bibinfo {pages}
		{712} (\bibinfo {year} {2017})}\BibitemShut {NoStop}%
	\bibitem [{\citenamefont {Zhong}\ \emph {et~al.}(2018)\citenamefont {Zhong},
		\citenamefont {Zhang},\ and\ \citenamefont {Li}}]{zhong_modeling_2018}%
	\BibitemOpen
	\bibfield  {author} {\bibinfo {author} {\bibfnamefont {L.}~\bibnamefont
			{Zhong}}, \bibinfo {author} {\bibfnamefont {Q.}~\bibnamefont {Zhang}}, \ and\
		\bibinfo {author} {\bibfnamefont {X.}~\bibnamefont {Li}},\ }\href@noop {}
	{\bibfield  {journal} {\bibinfo  {journal} {Scientific Reports}\ }\textbf
		{\bibinfo {volume} {8}},\ \bibinfo {pages} {2432} (\bibinfo {year}
		{2018})}\BibitemShut {NoStop}%
	\bibitem [{\citenamefont {Pretorius}\ \emph {et~al.}(2010)\citenamefont
		{Pretorius}, \citenamefont {Stover}, \citenamefont {Bollinger}, \citenamefont
		{Baca{\"e}r},\ and\ \citenamefont {Williams}}]{pretorius2010evaluating}%
	\BibitemOpen
	\bibfield  {author} {\bibinfo {author} {\bibfnamefont {C.}~\bibnamefont
			{Pretorius}}, \bibinfo {author} {\bibfnamefont {J.}~\bibnamefont {Stover}},
		\bibinfo {author} {\bibfnamefont {L.}~\bibnamefont {Bollinger}}, \bibinfo
		{author} {\bibfnamefont {N.}~\bibnamefont {Baca{\"e}r}}, \ and\ \bibinfo
		{author} {\bibfnamefont {B.}~\bibnamefont {Williams}},\ }\href@noop {}
	{\bibfield  {journal} {\bibinfo  {journal} {PloS one}\ }\textbf {\bibinfo
			{volume} {5}},\ \bibinfo {pages} {e13646} (\bibinfo {year}
		{2010})}\BibitemShut {NoStop}%
	\bibitem [{\citenamefont {Mushayabasa}\ \emph {et~al.}(2011)\citenamefont
		{Mushayabasa}, \citenamefont {Tchuenche}, \citenamefont {Bhunu},\ and\
		\citenamefont {Ngarakana-Gwasira}}]{mushayabasa2011modeling}%
	\BibitemOpen
	\bibfield  {author} {\bibinfo {author} {\bibfnamefont {S.}~\bibnamefont
			{Mushayabasa}}, \bibinfo {author} {\bibfnamefont {J.~M.}\ \bibnamefont
			{Tchuenche}}, \bibinfo {author} {\bibfnamefont {C.~P.}\ \bibnamefont
			{Bhunu}}, \ and\ \bibinfo {author} {\bibfnamefont {E.}~\bibnamefont
			{Ngarakana-Gwasira}},\ }\href@noop {} {\bibfield  {journal} {\bibinfo
			{journal} {BioSystems}\ }\textbf {\bibinfo {volume} {103}},\ \bibinfo {pages}
		{27} (\bibinfo {year} {2011})}\BibitemShut {NoStop}%
	\bibitem [{\citenamefont {Moreno}\ \emph {et~al.}(2002)\citenamefont {Moreno},
		\citenamefont {Pastor-Satorras},\ and\ \citenamefont
		{Vespignani}}]{moreno2002epidemic}%
	\BibitemOpen
	\bibfield  {author} {\bibinfo {author} {\bibfnamefont {Y.}~\bibnamefont
			{Moreno}}, \bibinfo {author} {\bibfnamefont {R.}~\bibnamefont
			{Pastor-Satorras}}, \ and\ \bibinfo {author} {\bibfnamefont {A.}~\bibnamefont
			{Vespignani}},\ }\href@noop {} {\bibfield  {journal} {\bibinfo  {journal}
			{The European Physical Journal B-Condensed Matter and Complex Systems}\
		}\textbf {\bibinfo {volume} {26}},\ \bibinfo {pages} {521} (\bibinfo {year}
		{2002})}\BibitemShut {NoStop}%
	\bibitem [{\citenamefont {Salath{\'e}}\ \emph {et~al.}(2010)\citenamefont
		{Salath{\'e}}, \citenamefont {Kazandjieva}, \citenamefont {Lee},
		\citenamefont {Levis}, \citenamefont {Feldman},\ and\ \citenamefont
		{Jones}}]{salathe2010high}%
	\BibitemOpen
	\bibfield  {author} {\bibinfo {author} {\bibfnamefont {M.}~\bibnamefont
			{Salath{\'e}}}, \bibinfo {author} {\bibfnamefont {M.}~\bibnamefont
			{Kazandjieva}}, \bibinfo {author} {\bibfnamefont {J.~W.}\ \bibnamefont
			{Lee}}, \bibinfo {author} {\bibfnamefont {P.}~\bibnamefont {Levis}}, \bibinfo
		{author} {\bibfnamefont {M.~W.}\ \bibnamefont {Feldman}}, \ and\ \bibinfo
		{author} {\bibfnamefont {J.~H.}\ \bibnamefont {Jones}},\ }\href@noop {}
	{\bibfield  {journal} {\bibinfo  {journal} {Proceedings of the National
				Academy of Sciences}\ }\textbf {\bibinfo {volume} {107}},\ \bibinfo {pages}
		{22020} (\bibinfo {year} {2010})}\BibitemShut {NoStop}%
	\bibitem [{\citenamefont {Barrat}\ \emph
		{et~al.}(2008{\natexlab{a}})\citenamefont {Barrat}, \citenamefont
		{Barthelemy},\ and\ \citenamefont {Vespignani}}]{Barrat2008b}%
	\BibitemOpen
	\bibfield  {author} {\bibinfo {author} {\bibfnamefont {A.}~\bibnamefont
			{Barrat}}, \bibinfo {author} {\bibfnamefont {M.}~\bibnamefont {Barthelemy}},
		\ and\ \bibinfo {author} {\bibfnamefont {A.}~\bibnamefont {Vespignani}},\
	}\href@noop {} {\emph {\bibinfo {title} {{Dynamical processes on complex
					networks}}}}\ (\bibinfo  {publisher} {Cambridge University Press},\ \bibinfo
	{year} {2008})\ p.\ \bibinfo {pages} {347}\BibitemShut {NoStop}%
	\bibitem [{\citenamefont {Gauvin}\ \emph {et~al.}(2013)\citenamefont {Gauvin},
		\citenamefont {Panisson}, \citenamefont {Cattuto},\ and\ \citenamefont
		{Barrat}}]{gauvin2013activity}%
	\BibitemOpen
	\bibfield  {author} {\bibinfo {author} {\bibfnamefont {L.}~\bibnamefont
			{Gauvin}}, \bibinfo {author} {\bibfnamefont {A.}~\bibnamefont {Panisson}},
		\bibinfo {author} {\bibfnamefont {C.}~\bibnamefont {Cattuto}}, \ and\
		\bibinfo {author} {\bibfnamefont {A.}~\bibnamefont {Barrat}},\ }\href@noop {}
	{\bibfield  {journal} {\bibinfo  {journal} {Scientific reports}\ }\textbf
		{\bibinfo {volume} {3}},\ \bibinfo {pages} {3099} (\bibinfo {year}
		{2013})}\BibitemShut {NoStop}%
	\bibitem [{\citenamefont {Barrat}\ \emph
		{et~al.}(2008{\natexlab{b}})\citenamefont {Barrat}, \citenamefont
		{Barthelemy},\ and\ \citenamefont {Vespignani}}]{barrat2008dynamical}%
	\BibitemOpen
	\bibfield  {author} {\bibinfo {author} {\bibfnamefont {A.}~\bibnamefont
			{Barrat}}, \bibinfo {author} {\bibfnamefont {M.}~\bibnamefont {Barthelemy}},
		\ and\ \bibinfo {author} {\bibfnamefont {A.}~\bibnamefont {Vespignani}},\
	}\href@noop {} {\emph {\bibinfo {title} {Dynamical processes on complex
				networks}}}\ (\bibinfo  {publisher} {Cambridge university press},\ \bibinfo
	{year} {2008})\BibitemShut {NoStop}%
	\bibitem [{\citenamefont {Holme}\ and\ \citenamefont
		{Litvak}(2017)}]{holme2017cost}%
	\BibitemOpen
	\bibfield  {author} {\bibinfo {author} {\bibfnamefont {P.}~\bibnamefont
			{Holme}}\ and\ \bibinfo {author} {\bibfnamefont {N.}~\bibnamefont {Litvak}},\
	}\href@noop {} {\bibfield  {journal} {\bibinfo  {journal} {PLoS computational
				biology}\ }\textbf {\bibinfo {volume} {13}},\ \bibinfo {pages} {e1005696}
		(\bibinfo {year} {2017})}\BibitemShut {NoStop}%
	\bibitem [{\citenamefont {Schiffman}(2003)}]{Schiffman2003}%
	\BibitemOpen
	\bibfield  {author} {\bibinfo {author} {\bibfnamefont {M.}~\bibnamefont
			{Schiffman}},\ }\href@noop {} {\bibfield  {journal} {\bibinfo  {journal}
			{Archives of pathology {\&} laboratory medicine}\ }\textbf {\bibinfo {volume}
			{127}},\ \bibinfo {pages} {930} (\bibinfo {year} {2003})}\BibitemShut
	{NoStop}%
	\bibitem [{\citenamefont {Cohen}\ \emph {et~al.}(2008)\citenamefont {Cohen},
		\citenamefont {Colijn},\ and\ \citenamefont {Murray}}]{Cohen16302}%
	\BibitemOpen
	\bibfield  {author} {\bibinfo {author} {\bibfnamefont {T.}~\bibnamefont
			{Cohen}}, \bibinfo {author} {\bibfnamefont {C.}~\bibnamefont {Colijn}}, \
		and\ \bibinfo {author} {\bibfnamefont {M.}~\bibnamefont {Murray}},\
	}\href@noop {} {\bibfield  {journal} {\bibinfo  {journal} {Proceedings of the
				National Academy of Sciences}\ }\textbf {\bibinfo {volume} {105}},\ \bibinfo
		{pages} {16302} (\bibinfo {year} {2008})}\BibitemShut {NoStop}%
	\bibitem [{\citenamefont {Poletto}\ \emph {et~al.}(2015)\citenamefont
		{Poletto}, \citenamefont {Meloni}, \citenamefont {Van~Metre}, \citenamefont
		{Colizza}, \citenamefont {Moreno},\ and\ \citenamefont
		{Vespignani}}]{poletto2015characterising}%
	\BibitemOpen
	\bibfield  {author} {\bibinfo {author} {\bibfnamefont {C.}~\bibnamefont
			{Poletto}}, \bibinfo {author} {\bibfnamefont {S.}~\bibnamefont {Meloni}},
		\bibinfo {author} {\bibfnamefont {A.}~\bibnamefont {Van~Metre}}, \bibinfo
		{author} {\bibfnamefont {V.}~\bibnamefont {Colizza}}, \bibinfo {author}
		{\bibfnamefont {Y.}~\bibnamefont {Moreno}}, \ and\ \bibinfo {author}
		{\bibfnamefont {A.}~\bibnamefont {Vespignani}},\ }\href@noop {} {\bibfield
		{journal} {\bibinfo  {journal} {Scientific reports}\ }\textbf {\bibinfo
			{volume} {5}},\ \bibinfo {pages} {7895} (\bibinfo {year} {2015})}\BibitemShut
	{NoStop}%
	\bibitem [{\citenamefont {Chen}\ \emph {et~al.}(2017)\citenamefont {Chen},
		\citenamefont {Ghanbarnejad},\ and\ \citenamefont
		{Brockmann}}]{chen2017fundamental}%
	\BibitemOpen
	\bibfield  {author} {\bibinfo {author} {\bibfnamefont {L.}~\bibnamefont
			{Chen}}, \bibinfo {author} {\bibfnamefont {F.}~\bibnamefont {Ghanbarnejad}},
		\ and\ \bibinfo {author} {\bibfnamefont {D.}~\bibnamefont {Brockmann}},\
	}\href@noop {} {\bibfield  {journal} {\bibinfo  {journal} {New Journal of
				Physics}\ }\textbf {\bibinfo {volume} {19}},\ \bibinfo {pages} {103041}
		(\bibinfo {year} {2017})}\BibitemShut {NoStop}%
	\bibitem [{\citenamefont {Pinotti}\ \emph {et~al.}(2018)\citenamefont
		{Pinotti}, \citenamefont {Fleury}, \citenamefont {Guillemot}, \citenamefont
		{B{\"o}elle},\ and\ \citenamefont {Poletto}}]{Pinotti2018}%
	\BibitemOpen
	\bibfield  {author} {\bibinfo {author} {\bibfnamefont {F.}~\bibnamefont
			{Pinotti}}, \bibinfo {author} {\bibfnamefont {{\'E}.}~\bibnamefont {Fleury}},
		\bibinfo {author} {\bibfnamefont {D.}~\bibnamefont {Guillemot}}, \bibinfo
		{author} {\bibfnamefont {P.-Y.}\ \bibnamefont {B{\"o}elle}}, \ and\ \bibinfo
		{author} {\bibfnamefont {C.}~\bibnamefont {Poletto}},\ }\href@noop {}
	{\bibfield  {journal} {\bibinfo  {journal} {bioRxiv}\ } (\bibinfo {year}
		{2018})}\BibitemShut {NoStop}%
	\bibitem [{\citenamefont {Rocha}\ \emph {et~al.}(2010)\citenamefont {Rocha},
		\citenamefont {Liljeros},\ and\ \citenamefont
		{Holme}}]{rocha2010information}%
	\BibitemOpen
	\bibfield  {author} {\bibinfo {author} {\bibfnamefont {L.~E.}\ \bibnamefont
			{Rocha}}, \bibinfo {author} {\bibfnamefont {F.}~\bibnamefont {Liljeros}}, \
		and\ \bibinfo {author} {\bibfnamefont {P.}~\bibnamefont {Holme}},\
	}\href@noop {} {\bibfield  {journal} {\bibinfo  {journal} {Proceedings of the
				National Academy of Sciences}\ }\textbf {\bibinfo {volume} {107}},\ \bibinfo
		{pages} {5706} (\bibinfo {year} {2010})}\BibitemShut {NoStop}%
	\bibitem [{\citenamefont {Rocha}\ \emph {et~al.}(2011)\citenamefont {Rocha},
		\citenamefont {Liljeros},\ and\ \citenamefont {Holme}}]{rocha2011simulated}%
	\BibitemOpen
	\bibfield  {author} {\bibinfo {author} {\bibfnamefont {L.~E.}\ \bibnamefont
			{Rocha}}, \bibinfo {author} {\bibfnamefont {F.}~\bibnamefont {Liljeros}}, \
		and\ \bibinfo {author} {\bibfnamefont {P.}~\bibnamefont {Holme}},\
	}\href@noop {} {\bibfield  {journal} {\bibinfo  {journal} {PLoS computational
				biology}\ }\textbf {\bibinfo {volume} {7}},\ \bibinfo {pages} {e1001109}
		(\bibinfo {year} {2011})}\BibitemShut {NoStop}%
	\bibitem [{\citenamefont {Holme}\ and\ \citenamefont
		{Saram{\"a}ki}(2013)}]{holme2013temporal}%
	\BibitemOpen
	\bibfield  {author} {\bibinfo {author} {\bibfnamefont {P.}~\bibnamefont
			{Holme}}\ and\ \bibinfo {author} {\bibfnamefont {J.}~\bibnamefont
			{Saram{\"a}ki}},\ }\href@noop {} {\emph {\bibinfo {title} {Temporal
				networks}}}\ (\bibinfo  {publisher} {Springer},\ \bibinfo {year}
	{2013})\BibitemShut {NoStop}%
	\bibitem [{\citenamefont {Kretzschmar}\ \emph {et~al.}(1996)\citenamefont
		{Kretzschmar}, \citenamefont {Duynhoven},\ and\ \citenamefont
		{Severijnen}}]{kretzschmar_modeling_1996}%
	\BibitemOpen
	\bibfield  {author} {\bibinfo {author} {\bibfnamefont {M.~E.}\ \bibnamefont
			{Kretzschmar}}, \bibinfo {author} {\bibfnamefont {Y.~T.~v.}\ \bibnamefont
			{Duynhoven}}, \ and\ \bibinfo {author} {\bibfnamefont {A.~J.~d.}\
			\bibnamefont {Severijnen}},\ }\href@noop {} {\bibfield  {journal} {\bibinfo
			{journal} {American journal of epidemiology}\ }\textbf {\bibinfo {volume}
			{144}},\ \bibinfo {pages} {306} (\bibinfo {year} {1996})}\BibitemShut
	{NoStop}%
	\bibitem [{\citenamefont {Bekker}\ \emph {et~al.}(2015)\citenamefont {Bekker},
		\citenamefont {Johnson}, \citenamefont {Cowan}, \citenamefont {Overs},
		\citenamefont {Besada}, \citenamefont {Hillier},\ and\ \citenamefont
		{Cates}}]{bekker_combination_2015}%
	\BibitemOpen
	\bibfield  {author} {\bibinfo {author} {\bibfnamefont {L.-G.}\ \bibnamefont
			{Bekker}}, \bibinfo {author} {\bibfnamefont {L.}~\bibnamefont {Johnson}},
		\bibinfo {author} {\bibfnamefont {F.}~\bibnamefont {Cowan}}, \bibinfo
		{author} {\bibfnamefont {C.}~\bibnamefont {Overs}}, \bibinfo {author}
		{\bibfnamefont {D.}~\bibnamefont {Besada}}, \bibinfo {author} {\bibfnamefont
			{S.}~\bibnamefont {Hillier}}, \ and\ \bibinfo {author} {\bibfnamefont
			{W.}~\bibnamefont {Cates}},\ }\href@noop {} {\bibfield  {journal} {\bibinfo
			{journal} {The Lancet}\ }\textbf {\bibinfo {volume} {385}},\ \bibinfo {pages}
		{72} (\bibinfo {year} {2015})}\BibitemShut {NoStop}%
	\bibitem [{\citenamefont {Chesson}\ and\ \citenamefont
		{Pinkerton}(2000)}]{chesson_sexually_2000}%
	\BibitemOpen
	\bibfield  {author} {\bibinfo {author} {\bibfnamefont {H.~W.}\ \bibnamefont
			{Chesson}}\ and\ \bibinfo {author} {\bibfnamefont {S.~D.}\ \bibnamefont
			{Pinkerton}},\ }\href@noop {} {\bibfield  {journal} {\bibinfo  {journal}
			{JAIDS Journal of Acquired Immune Deficiency Syndromes}\ }\textbf {\bibinfo
			{volume} {24}},\ \bibinfo {pages} {48} (\bibinfo {year} {2000})}\BibitemShut
	{NoStop}%
	\bibitem [{\citenamefont {Mukandavire}\ \emph {et~al.}(2016)\citenamefont
		{Mukandavire}, \citenamefont {Mitchell},\ and\ \citenamefont
		{Vickerman}}]{mukandavire_comparing_2016}%
	\BibitemOpen
	\bibfield  {author} {\bibinfo {author} {\bibfnamefont {Z.}~\bibnamefont
			{Mukandavire}}, \bibinfo {author} {\bibfnamefont {K.~M.}\ \bibnamefont
			{Mitchell}}, \ and\ \bibinfo {author} {\bibfnamefont {P.}~\bibnamefont
			{Vickerman}},\ }\href@noop {} {\bibfield  {journal} {\bibinfo  {journal}
			{Epidemics}\ }\textbf {\bibinfo {volume} {14}},\ \bibinfo {pages} {62}
		(\bibinfo {year} {2016})}\BibitemShut {NoStop}%
	\bibitem [{\citenamefont {MacFadden}\ \emph {et~al.}(2016)\citenamefont
		{MacFadden}, \citenamefont {Tan},\ and\ \citenamefont
		{Mishra}}]{macfadden_optimizing_2016}%
	\BibitemOpen
	\bibfield  {author} {\bibinfo {author} {\bibfnamefont {D.~R.}\ \bibnamefont
			{MacFadden}}, \bibinfo {author} {\bibfnamefont {D.~H.}\ \bibnamefont {Tan}},
		\ and\ \bibinfo {author} {\bibfnamefont {S.}~\bibnamefont {Mishra}},\
	}\href@noop {} {\bibfield  {journal} {\bibinfo  {journal} {Journal of the
				International AIDS Society}\ }\textbf {\bibinfo {volume} {19}} (\bibinfo
		{year} {2016})}\BibitemShut {NoStop}%
	\bibitem [{\citenamefont {Tripathi}\ \emph {et~al.}(2007)\citenamefont
		{Tripathi}, \citenamefont {Naresh},\ and\ \citenamefont
		{Sharma}}]{tripathi2007modeling}%
	\BibitemOpen
	\bibfield  {author} {\bibinfo {author} {\bibfnamefont {A.}~\bibnamefont
			{Tripathi}}, \bibinfo {author} {\bibfnamefont {R.}~\bibnamefont {Naresh}}, \
		and\ \bibinfo {author} {\bibfnamefont {D.}~\bibnamefont {Sharma}},\
	}\href@noop {} {\bibfield  {journal} {\bibinfo  {journal} {Applied
				mathematics and computation}\ }\textbf {\bibinfo {volume} {184}},\ \bibinfo
		{pages} {1053} (\bibinfo {year} {2007})}\BibitemShut {NoStop}%
	\bibitem [{\citenamefont {Stehl{\'e}}\ \emph {et~al.}(2011)\citenamefont
		{Stehl{\'e}}, \citenamefont {Voirin}, \citenamefont {Barrat}, \citenamefont
		{Cattuto}, \citenamefont {Colizza}, \citenamefont {Isella}, \citenamefont
		{R{\'e}gis}, \citenamefont {Pinton}, \citenamefont {Khanafer}, \citenamefont
		{Van~den Broeck},\ and\ \citenamefont {Vanhems}}]{Stehle2011}%
	\BibitemOpen
	\bibfield  {author} {\bibinfo {author} {\bibfnamefont {J.}~\bibnamefont
			{Stehl{\'e}}}, \bibinfo {author} {\bibfnamefont {N.}~\bibnamefont {Voirin}},
		\bibinfo {author} {\bibfnamefont {A.}~\bibnamefont {Barrat}}, \bibinfo
		{author} {\bibfnamefont {C.}~\bibnamefont {Cattuto}}, \bibinfo {author}
		{\bibfnamefont {V.}~\bibnamefont {Colizza}}, \bibinfo {author} {\bibfnamefont
			{L.}~\bibnamefont {Isella}}, \bibinfo {author} {\bibfnamefont
			{C.}~\bibnamefont {R{\'e}gis}}, \bibinfo {author} {\bibfnamefont {J.-F.}\
			\bibnamefont {Pinton}}, \bibinfo {author} {\bibfnamefont {N.}~\bibnamefont
			{Khanafer}}, \bibinfo {author} {\bibfnamefont {W.}~\bibnamefont {Van~den
				Broeck}}, \ and\ \bibinfo {author} {\bibfnamefont {P.}~\bibnamefont
			{Vanhems}},\ }\href@noop {} {\bibfield  {journal} {\bibinfo  {journal} {BMC
				Medicine"}\ }\textbf {\bibinfo {volume} {9}},\ \bibinfo {pages} {87}
		(\bibinfo {year} {2011})}\BibitemShut {NoStop}%
	\bibitem [{\citenamefont {Dimitrov}\ \emph {et~al.}(2016)\citenamefont
		{Dimitrov}, \citenamefont {M{\^a}sse},\ and\ \citenamefont
		{Donnell}}]{dimitrov2016prep}%
	\BibitemOpen
	\bibfield  {author} {\bibinfo {author} {\bibfnamefont {D.~T.}\ \bibnamefont
			{Dimitrov}}, \bibinfo {author} {\bibfnamefont {B.~R.}\ \bibnamefont
			{M{\^a}sse}}, \ and\ \bibinfo {author} {\bibfnamefont {D.}~\bibnamefont
			{Donnell}},\ }\href@noop {} {\bibfield  {journal} {\bibinfo  {journal}
			{Journal of acquired immune deficiency syndromes (1999)}\ }\textbf {\bibinfo
			{volume} {72}},\ \bibinfo {pages} {444} (\bibinfo {year} {2016})}\BibitemShut
	{NoStop}%
	\bibitem [{\citenamefont {Cohen}\ \emph {et~al.}(2003)\citenamefont {Cohen},
		\citenamefont {Havlin},\ and\ \citenamefont {ben Avraham}}]{Cohen2003}%
	\BibitemOpen
	\bibfield  {author} {\bibinfo {author} {\bibfnamefont {R.}~\bibnamefont
			{Cohen}}, \bibinfo {author} {\bibfnamefont {S.}~\bibnamefont {Havlin}}, \
		and\ \bibinfo {author} {\bibfnamefont {D.}~\bibnamefont {ben Avraham}},\
	}\href@noop {} {\bibfield  {journal} {\bibinfo  {journal} {Phys. Rev. Lett.}\
		}\textbf {\bibinfo {volume} {91}},\ \bibinfo {pages} {247901} (\bibinfo
		{year} {2003})}\BibitemShut {NoStop}%
\end{thebibliography}
\end{document}